\documentstyle[12pt,a41,epsfig]{article}
\newcommand{\bq}{\begin{equation}}
\newcommand{\eq}{\end{equation}}

\newcommand{\lsim}{\raisebox{-0.07cm}{$\, \stackrel{<}{{\scriptstyle
\sim}}\, $}}
\newcommand{\gsim}{\raisebox{-0.07cm}{$\, \stackrel{>}{{\scriptstyle
\sim}}\, $}}

\newcommand\GeV{\,\mbox{GeV}}
\newcommand\TeV{\,\mbox{TeV}}

\newcommand\CALL{{\cal L}}
\newcommand\KG{\kappa_G}

\newcommand\LG{\lambda_G}

\newcommand\PD{\not p}
\newcommand\MP{M^2_{\Phi}}
\newcommand\B{\beta}
\newcommand\XL{\log \left| \frac{1 + \B}{1 - \B} \right |}
\newcommand\SM{\frac{\hat{s}}{M_{\Phi}^2}}
\newcommand\SMM{\frac{\hat{s}^2}{M_{\Phi}^4}}
\newcommand\SMMM{\frac{\hat{s}^3}{M_{\Phi}^6}}
\newcommand\SMMMM{\frac{\hat{s}^4}{M_{\Phi}^8}}

\newcommand\CB{\beta^2 \cos^2 \theta}
\newcommand\CBB{\beta^4 \cos^4 \theta}

\newcommand\uh{\hat{u}}
\newcommand\thh{\hat{t}}
\newcommand\sh{\hat{s}}
\newcommand\ea{\varepsilon_1}
\newcommand\eb{\varepsilon_2}
\newcommand\Ea{\epsilon_1}
\newcommand\Eb{\epsilon_2}
\newcommand\ch{{\rm{ch}}}
\newcommand\shh{{\rm{sh}}}
\newcommand\thhh{{\rm{th}}}
\newcommand\st{\sin^2 \theta}
\newcommand\BR{\left(1 - \CB \right)}

\begin{document}
\sloppy
\thispagestyle{empty}
\begin{flushleft}
DESY 96--174 \\
October  1996
\end{flushleft}

\mbox{}
\vspace*{\fill}
\begin{center}
{\LARGE\bf Leptoquark Pair Production in Hadronic} \\

\vspace{2mm}
{\LARGE\bf Interactions}\\

\vspace{2em}
\large
Johannes Bl\"umlein$^1$, Edward Boos$^{1,2}$, and
Alexander Kryukov$^{1,2}$
\\
\vspace{2em}
{\it $^1$DESY --Zeuthen,}
 \\
{\it Platanenallee 6,
D--15735 Zeuthen, Germany}\\
{\it $^2$Institute of Nuclear Physics, Moscow State University,}\\
{\it RU--119899 Moscow, Russia}\\
\end{center}
\vspace*{\fill}
\begin{abstract}
\noindent
The scalar and vector leptoquark pair production cross sections in
hadronic collisions are calculated. In a model independent analysis
we consider the  most general $C$ and $P$ conserving couplings of
gluons to both scalar and vector leptoquarks described by an effective
low--energy  Lagangian which obeys $SU(3)_c$ invariance. Analytrical
expressions are derived for the differential and integral scattering
cross sections including the case of anomalous vector leptoquark
couplings, $\kappa_G$ and $\lambda_G$, to the gluon field. Numerical
predictions are given for the kinematic range of the TEVATRON and LHC.
The pair production cross sections are also calculated for the resolved
photon contributions to $ep \rightarrow e \overline{\Phi} \Phi X $  at
HERA and LEP~$\otimes$~LHC, and for the process $\gamma \gamma
\rightarrow \Phi \overline {\Phi}X$ at possible future $e^+e^-$ linear
colliders and $\gamma \gamma$ colliders. Estimates of the search potential
for scalar and vector leptoquarks at present and future high energy
colliders are given.
\end{abstract}
\vspace*{\fill}
\newpage
%
\section{Introduction}
\label{sect1}
\noindent
In many extensions of the Standard Model new bosons are predicted which
carry both lepton and baryon number~\cite{LQ1}. There is indeed a
close relation between these quantum numbers in the Standard
Model since the triangle anomalies are cancelled by the requirement
\begin{equation}
\label{ein1}
\sum_n Q_n^2 \left ( Q_L - Q_R \right )_n = 0
\end{equation}
for each fermion family,
which renders the theory renormalizable. Here $Q_n, Q_{Ln}$, and $Q_{Rn}$
denote the electromagnetic, left-, and right-handed neutral current
charges, respectively.
Leptoquark states emerge
naturally as the gauge bosons in Grand Unified Theories~\cite{LQ2}.
In these scenarios
their couplings are not baryon number conserving and
their masses are  situated in the range $M_{\Phi} \gsim 10^{15}
\GeV$.
On the other hand, leptoquarks with B- and L-conserving couplings may
exist in
the mass range accessible at high energy colliders. These states will
be considered in the present paper. They may be either
sought  through virtual effects
in low energy processes, from which already
severe constraints on
their fermionic couplings were derived~\cite{LEUR}, or searched for at
high energy colliders as
at LEP~\cite{LQLEP}, HERA~\cite{LQHERA}, and
TEVATRON~\cite{LQTEVA}. These searches
constrained further the allowed mass and
coupling  ranges being limited by other experiments~\cite{LQEARLY}
previously. At present the most stringent constraints come from the
TEVATRON  and exclude leptoquarks associated with the first and second
family with masses in the range
$M_{\Phi} \lsim O(100) \GeV$~\cite{PDG}. For the third generation scalar
leptoquarks currently the mass range $M_{\Phi} < 45 \GeV$ is excluded.

In the case of single leptoquark production the scattering cross sections
are always proportional to one
of the fermionic couplings~\cite{SINGLE},
$\lambda_{lq}$, which are
constrained to be rather small from low energy
processes up to masses of
$M_{\Phi} \sim O(1 \TeV)$~\footnote{For even higher
masses the fermion couplings of leptoquarks can still be of the
order $\lambda_{lq} \sim e$. Single production at proton colliders
has been studied
in~\cite{ZERWAS} recently.}. Thus the search limits {\it always}
lead  to combined bounds on the couplings $\lambda_{lq}$ and
the leptoquark masses.
For leptoquark pair production, however, the small size
    of the fermionic
couplings does not severely constrain
the scattering cross sections due to the finite bosonic contributions.
The strength of the leptoquark couplings to the gauge bosons,
$\gamma, g, W^{\pm}$, and $Z^0$, is  determined by the coupling constant
of the respective gauge field up to eventual anomalous contributions
in the case of vector leptoquarks,
which are described by the parameters $\kappa_A$ and $\lambda_A$,
with
$A = \gamma, g, W^{\pm}$, and $Z^0$. As will be shown for the hadronic
contributions to the different
scattering cross sections, a combination of
anomalous couplings $(\kappa_G, \lambda_G)$ exists
for which they become minimal. Therefore one may constrain the
allowed mass ranges for leptoquarks {\it directly}. As will be shown,
the corresponding minimizing
couplings are in general {\it not} those of the Yang-Mills type
or the minimal vector boson couplings.

In the present paper the cross sections for scalar and vector leptoquark
pair production in hadronic interactions are calculated.
They apply for
$p\overline{p}$ and $pp$ scattering at the
TEVATRON and LHC, respectively,
but also for hadronic interactions at $ep$, $e^+e^-$, and
$\gamma \gamma$ colliders. In the latter cases they emerge as the
resolved photon contributions which accompany the respective direct
photon processes~\cite{BBP}--\cite{BB94}.

The paper is organized as follows.
In section~2 the basic notations are introduced. Section~3 contains the
derivation of the partonic scattering cross sections for scalar
and vector leptoquark pair production including the anomalous
couplings $\kappa_A$ and $\lambda_A$. Here it is also shown how the
scattering cross sections for the scalar and the vector case, in
the absence of anomalous couplings, can be obtained using factorization
relations of the amplitude. The hadronic production cross sections are
derived in section~4. For $ep$, $e^+e^-$, and $\gamma \gamma$ scattering
the direct photon contributions
calculated previously~\cite{BBP}--\cite{BB94}
are added to obtain a complete description. The numerical results
are presented in section~5~\footnote{The code for
the different
processes can be obtained on request from {\sf blumlein@ifh.de}.
A detailed description of the code {\sf LQPAIR 1.0} is given
in~\cite{LQPAIR}.}
and section~6 contains the conclusions. The Feynman rules used in the
present calculation are summarized in Appendix~A. Appendix~B contains
the coefficients which
describe the differential and integrated pair
production cross sections for vector leptoquarks in the presence of
anomalous couplings.
\section{Basic Notations}
\label{sect2A}
Following earlier investigations~\cite{BBP}--\cite{BB94},\cite{BR}
we consider the class
of leptoquarks introduced in~\cite{BRW}. The fermionic couplings
of these states are dimensionless, baryon and lepton number conserving,
family--diagonal, and $SU(3)_c \times SU(2)_L \times U(1)_Y$ invariant.
These
leptoquarks are color triplets.
As outlined in ref.~\cite{BR}, these conditions do widely induce also
the couplings to the gauge bosons of the Standard Model.
For the scalar states their
bosonic couplings are determined completely. In the case of
vector leptoquarks, the
Yang--Mills type couplings may be supplemented by anomalous couplings
which are specified by two parameters $\kappa_A$ and $\lambda_A$.
The couplings $\kappa_A$ and $\lambda_A$,
corresponding to different gauge fields,
are not generally related.
The hadronic processes depend  on
the parameters
$\KG$ and $\LG$ only. Since
most of the fermionic couplings $\lambda_{lq}$
of the
leptoquarks are bounded to  be very small in the mass range up to
${\cal O}(1 \TeV)$~\cite{LEUR},
we will neglect their contribution
in the following and  consider pair production through
bosonic couplings only\footnote{
The hadronic pair production cross sections
for leptoquarks of different flavor $\Phi_1 \overline{\Phi}_2$ or
$\Phi_1 \Phi_2$ is possible through quark or quark--antiquark scattering
and depends on the  fermionic couplings as $\propto \lambda_{lq}^4$.}.

The effective
Lagrangian describing the interaction of the scalar and vector
leptoquarks with gluons is given by
\begin{equation}
\label{eqLA}
\CALL =
\CALL_S^g + \CALL_V^g,
\end{equation}
where
\begin{equation}
\label{eqLAS}
\CALL_S^g
 = \sum_{scalars} \left [
\left (D^{\mu}_{ij} \Phi^j \right )^{\dagger}
                                 \left (D_{\mu}^{ik} \Phi_k \right )
 - M_S^2 \Phi^{i \dagger} \Phi_i \right ],
\end{equation}
\begin{equation}
\label{eqLAV}
\CALL_V^g
= \sum_{vectors} \left \{ -\frac{1}{2} G^{i \dagger}_{\mu \nu}
G^{\mu \nu}_i + M_V^2 \Phi_{\mu}^{i \dagger} \Phi^{\mu}_i \right.
- \left.
ig_s \left [ (1 - \KG)
\Phi_{\mu}^{i \dagger}
t^a_{ij}
\Phi_{\nu}^j
{\cal G}^{\mu \nu}_a
+ \frac{\LG}{M_V^2} G^{i\dagger}_{\sigma \mu}
t^a_{ij}
G_{\nu}^{j \mu} {\cal G}^{\nu \sigma}_a \right ] \right \}.
\end{equation}
Here,
$g_s$ denotes the  strong coupling
constant, $t_a$ are the generators of $SU(3)_c$,
$M_S$ and $M_V$
are
the leptoquark masses, and $\KG$ and
$\LG$ are the anomalous couplings.
The field strength tensors of the  gluon and vector
leptoquark fields are
\begin{eqnarray}
{\cal G}_{\mu \nu}^a  &=& \partial_{\mu} {\cal A}_{\nu}^a
 - \partial_{\nu}
{\cal A}_{\mu}^a + g_s f^{abc} {\cal A}_{\mu b} {\cal A}_{\nu c},
 \nonumber\\
G_{\mu \nu}^{i}
 &=& D_{\mu}^{ik}
 \Phi_{\nu k} - D_{\nu}^{ik} \Phi_{\mu k},
\end{eqnarray}
with the covariant derivative given by
\begin{equation}
D_{\mu}^{ij} = \partial_{\mu} \delta^{ij} - i g_s
t_a^{ij}
 {\cal A}^a_{\mu}.
\end{equation}

The parameters $\KG$ and $\LG$ are assumed to be
real. They are related
to the anomalous 'magnetic' moment
$\mu_V$ and 'electric' quadrupole moment $q_V$
of the leptoquarks in the color field
\begin{eqnarray}
\mu_{V,G} = \frac{g_s}{2 M_{V}} \left ( 2 -
\kappa_G + \lambda_G \right ), \nonumber\\
q_{V,G} = - \frac{g_s}
{M_{V}^2} \left (1 - \kappa_{G} - \lambda_{G} \right ).
\end{eqnarray}
Since we wish to keep the analysis as model independent as possible
we assume that these quantities are   independent.
At present there are no direct bounds on the parameters $\kappa_G$
and $\lambda_G$. Below we will consider the range of
$|\kappa_G|,|\lambda_G| \leq 1$. The hadronic production cross sections
are found to vary significantly for parameters in this range.
As will be shown,  searches at the TEVATRON will be able to
constrain this range further.
The above choice covers
both the cases of Yang--Mills type couplings, $\kappa_G = \lambda_G
= 0$, and the minimal vector couplings, $\kappa_G =1, \lambda_G = 0$.

The Feynman rules relevant for the processes studied in the present
paper  are summarized in
Appendix~A.
\section{Partonic Cross Sections}
\label{PARCS}
\noindent
Before we study the pair production cross sections for leptoquarks
at different colliders we  present the partonic cross sections.
The diagrams of the contributing subprocesses
$gg \rightarrow \Phi \overline{\Phi}$ and
$q \overline{q} \rightarrow \Phi \overline{\Phi}$ are shown in
figure~1 and~2.
Let us first consider the case of vanishing anomalous couplings, i.e.
the simplified situation in which the production cross sections depend
on the gauge coupling {\it only}.

In non-Abelian gauge theories the amplitudes for a series of
$2 \rightarrow 2$ scattering processes can be factorized into a
group and a Lorentz part~\cite{ZHU}. As will be shown
this applies to ${\cal M}(gg \rightarrow \Phi_S \overline{\Phi}_S)$ and
${\cal M}(gg \rightarrow \Phi_V \overline{\Phi}_V)$ for the special case
of vanishing anomalous vector couplings $\KG = \LG \equiv 0$.
This representation also yields
a particularly simple result for the differential cross sections
in comparison with expressions obtained otherwise~\cite{SCAL1,VECCOR}.
The former case
has been dealt with in ref.~\cite{ZHU} using this method, however,
the scattering cross section obtained disagrees with other
results~\cite{SCAL1}.
Therefore we will recalculate the
cross sections for both cases showing the factorization of the
group factor
and the Lorentz part in detail  before deriving the more
general result
for finite anomalous couplings $\KG$ and $\LG$.

\subsection{Factorization Relations}
\label{sectXX}

We use a physical gauge for the gluon fields. The gluon polarization
vectors $\varepsilon_g$ obey $\varepsilon_g.p_g = 0$ and
$\sum_{\lambda} \varepsilon^{\mu}_g(\lambda)
 \varepsilon^{\nu*}_g(\lambda) = - g^{\mu \nu} + p_g^{\mu} p_g^{\nu}
/p_g.p_g$.
The matrix elements
for the above processes can be written as
\begin{equation}
{\cal M}^{S,V} = \sum_p g_s^2
G_p       \frac{T^{S,V}_p}{C_p},
\end{equation}
with
$p = s,t,u$ the channel index,
 $C_s = s, C_t \equiv \hat{t} =
t - \MP$, and $C_u \equiv \hat{u} =
u - \MP$. The numerators
in the matrix element have been written in terms of the respective
group factors $G_p$ and the Lorentz parts $T^{S,V}_p$.
The following relations are valid.
\begin{eqnarray}
\label{EQ9}
C_s + C_u + C_t &=& 0, \\
\label{EQ10}
G_t - G_u       &=& G_s, \\
\label{EQ11}
\hat{T}_t - \hat{T}_u &=& T_s + \Delta .
\end{eqnarray}
Since $G_t = (t^a t^b)_{ij}$,
      $G_u = (t^b t^a)_{ij}$, and
$G_s =  i f_{abc} t^c_{ij}$, eq.~(\ref{EQ10})
is the commutation relation of the generators of $SU(3)_c$.
The Lorentz terms $\hat{T}_{t,u}$ consist out of the $t$ and $u$~channel
terms supplemented by the parts of the sea--gull diagram (see
figure~1)
corresponding
to the group factor $G_t$ and $G_u$, respectively.
The contributions to the Lorentz parts $T_s, \hat{T}_t$ and $\hat{T}_u$
are given  in table~1 for the scalar and vector case.
One finds that
\begin{equation}
\label{eqdelt}
\Delta = p_1.\varepsilon(p_1) A +  p_2.\varepsilon(p_2) B
\equiv 0,
\end{equation}
with $\varepsilon(p_i)$ the  gluon polarization vectors.
These relations result into
\begin{equation}
\label{MATEL}
\left | {\cal M} \right |^2 = |G|^2 \left |{\cal M}_A \right |^2
= |G|^2 g_s^2
\left | \frac{\hat{T}_t}{C_t}
                                      + \frac{\hat{T}_u}{C_u} \right |^2.
\end{equation}
${\cal M}_A$ has thus
the form of an Abelian amplitude. Since the leptoquarks
described by~(\ref{eqLA}) are color triplets or antitriplets
one obtains
\begin{equation}
\left |G \right|^2 = \left |\frac{C_u G_t + C_t G_u}{C_s} \right |^2 =
\frac{16}{3} \frac{\uh^2 + \thh^2}{s^2}
- \frac{4}{3} \frac{\uh \thh}{s^2} =
4 \left ( \frac{4}{3} - 3 \frac{\hat{u} \hat{t}}
                     {s^2} \right ).
\end{equation}

\vspace{5mm}
\begin{center}
\begin{tabular}{|c||l|l|}
\hline \hline
\multicolumn{1}{|c||}{          }
  & \multicolumn{1}{c|}{{\sf Scalar Field}}
  & \multicolumn{1}{c|}{{\sf Vector Field, $\kappa_G = \lambda_G =0$
}}\\
\hline \hline
$T_s$ &$ - (q_1 - q_2).\ea (2 p_1 + p_2).\eb$
 &
$\ea.\eb \Ea.\Eb (\uh - \thh)$ \\
 &$      + (q_1 - q_2).\eb (2 p_2 + p_1).\ea$ &
$+ 4 \ea.\eb (-p_1.\Ea p_2.\Eb + p_1.\Eb p_2.\Ea)$ \\
 &$ + (\uh - \thh) \ea.\eb$
        &$ -4 \ea.\Ea  (p_1.\Eb q_1.\eb + p_1.\Eb q_2.\eb
                    +   p_2.\Eb q_1.\eb + p_2.\Eb q_2.\eb)$ \\
      &  &$ +4 \ea.\Eb  (p_1.\Ea q_1.\eb + p_1.\Ea q_2.\eb
                    +   p_2.\Ea q_1.\eb + p_2.\Ea q_2.\eb)$\\
      &  &$ +4 \eb.\Ea (p_1.\Eb q_1.\ea + p_1.\Eb q_2.\ea
                    +   p_2.\Eb q_1.\ea + p_2.\Eb q_2.\ea )$\\
      &  &$ -4 \eb.\Eb (p_1.\Ea q_1.\ea + p_1.\Ea q_2.\ea
                      + p_2.\Ea q_1.\ea + p_2.\Ea q_2.\ea)$ \\
      &  &$ +4 \Ea.\Eb (-q_1.\ea q_2.\eb + q_1.\eb q_2.\ea )$\\
\hline
$\hat{T}_t$ &
$(2 q_1 - p_1).\ea (p_2 - 2 q_2).\eb - \thh \ea.\eb$
  & $  - \ea.\eb \Ea.\Eb \thh$ \\
            &   & $  + 2 ( \ea.\Ea \eb.\Eb \uh
+ \ea.\Eb \eb.\Ea \thh)$\\
            &   & $  -4 ( \ea.\eb p_1.\Ea p_2.\Eb
                        + \Ea.\Eb q_1.\ea q_2.\eb )$ \\
            &   & $  +4 ( \ea.\Eb p_1.\Ea q_2.\eb
                        + \eb.\Ea p_2.\Eb q_1.\ea )$  \\
   & & $    -4 \ea.\Ea (p_1.\Eb q_2.\eb + p_2.\Eb q_1.\eb
                   +   p_2.\Eb q_2.\eb )$\\
   & & $    -4 \eb.\Eb (p_1.\Ea q_1.\ea + p_1.\Ea q_2.\ea
                     +  p_2.\Ea q_1.\ea )$\\
\hline
$\hat{T}_u$ &
$(2 q_1 - p_2).\ea (p_1 - 2 q_2).\eb - \uh \ea.\eb$
  & $  - \ea.\eb \Ea.\Eb \uh$ \\
            &   & $  + 2 ( \ea.\Ea \eb.\Eb \uh
+ \ea.\Eb \eb.\Ea \thh)$\\
            &   & $  -4 ( \ea.\eb p_1.\Eb p_2.\Ea
                        + \Ea.\Eb q_1.\eb q_2.\ea )$ \\
            &   & $  +4 ( \ea.\Ea p_1.\Eb q_1.\eb
                        + \eb.\Eb p_2.\Ea q_2.\eb )$  \\
   & & $    -4 \ea.\Eb (p_1.\Ea q_1.\eb + p_2.\Ea q_1.\eb
                   +   p_2.\Ea q_2.\eb )$\\
   & & $    -4 \eb.\Ea (p_1.\Eb q_1.\ea + p_1.\Eb q_2.\ea
                     +  p_2.\Eb q_2.\ea )$\\
\hline \hline
\end{tabular}
\label{T1}
\normalsize
\end{center}
\small

\vspace*{2mm}
\noindent
{\sf Table~1:~$stu$ contributions to the Lorentz part of
the scattering amplitudes
$gg \rightarrow \Phi \overline{\Phi}$. The polarization vectors of
the vector field are denoted by $\epsilon_{1,2}$.}
\\
\normalsize

\vspace{3mm}
\noindent
The Lorentz parts of the amplitude yield
\begin{equation}
\label{eqggss}
\left |{\cal M}_A^S \right|^2
= 8 g_s^4 \left [1 - 2 \frac{s M_S^2}{\thh \uh}
                   + 2 \left (\frac{s M_S^2}{\thh \uh} \right)^2
          \right ],
\end{equation}
and
\begin{equation}
\label{eqggvv}
\left |{\cal M}_A^V \right|^2
= 8 g_s^4 \left [3 - 2 \left (2 - \frac{s^2}{\thh \uh} \right )
\frac{s^2}{\thh \uh} - 6 \left ( 1 - \frac{M^2_V s}{\thh \uh} \right )
\frac{M^2_V s}{\thh \uh}     \right ].
\end{equation}

Comparing (\ref{MATEL}) with other results obtained in
earlier calculations,  we agree with the scattering cross sections
for scalar pair production derived in \cite{SCAL1}
but disagree with those found
in~\cite{ZHU,SCAL2,SCAL4}\footnote{
We checked that the difference in $\hat{\sigma}_{S\overline{S}}^{gg}$
in ref.~\cite{SCAL2} could be explained by leaving out the ghost term
in the Feynman gauge.
The result obtained in
ref.~\cite{SCAL5} agrees with that given in \cite{SCAL2} and
was used to derive numerical
results. Later the same authors revised this expression, 
see~\cite{MMS},
and agree with eq.~(\ref{eq4a}).
Since ref.~[11] in \cite{SCAL3} refers to two {\it different}
expressions for $\sigma(gg \rightarrow \Phi_s \overline{\Phi}_s)$ it
remains
unclear on which relation the numerical calculation presented
was based.}.
Our result for pair production of vector color
triplets agrees with that
given in \cite{VECCOR,SCAL4}\footnote{A numerical illustration of this
relation has been given in \cite{EBO94} for the gluonic contributions
recently.}.

\subsection{Scalar Leptoquarks}
\label{sect2B}
\noindent
The differential and integral pair production cross sections for
$gg$ and
$q\overline{q}$  scattering are
\begin{eqnarray}
\label{eq3a}
\frac{d \hat{\sigma}_{S\overline{S}}^{gg}}{d \cos \theta}
&=& \frac{\pi \alpha_s^2}{6  \hat{s}} \beta
\left \{
\frac{1}{32} \left [ 25 + 9 \CB - 18 \beta^2 \right ] \right.
\nonumber\\
&-& \left.
\frac{1}{16} \frac{( 25 -34 \beta^2 + 9 \beta^4)}{1 - \CB}
+  \frac{(1 - \beta^2)^2}{(1 - \CB)^2} \right \},
\end{eqnarray}
\begin{equation}
\label{eq4a}
\hat{\sigma}_{S\overline{S}}^{gg} =
\frac{\pi \alpha_s^2}{96 \hat{s}} \left \{ \beta
\left(41 - 31 \beta^2 \right)
 - \left (17 - 18 \beta^2 + \beta^4 \right)
 \log \left| \frac{1 + \beta}
{1 -  \beta} \right | \right \},
\end{equation}
and
\begin{equation}
\label{eq1a}
\frac{d \hat{\sigma}_{S\overline{S}}^{q\overline{q}}}{d \cos \theta}
= \frac{\pi \alpha_s^2}{18 \sh} \beta^3 \sin^2 \theta,
\end{equation}
\begin{equation}
\label{eq2a}
\hat{\sigma}_{S\overline{S}}^{q\overline{q}}
= \frac{2 \pi \alpha_s^2}{27 \sh} \beta^3,
\end{equation}
with $\alpha_s = g^2_s/4 \pi$,
$\beta = \sqrt{1 - 4 M_{\Phi}^2/\hat{s}}$, $\hat{s}^{1/2}$ the cms
energy and $\theta$ the leptoquark scattering angle in the parton--parton
cms. All quark flavors have been dealt with
as massless\footnote{This is a sufficient approximation in the mass
and energy range
$M_{\Phi} > 40 \GeV$ and $\sqrt{S} \ge 300 \GeV$ considered in
the present paper.}.
Eqs. (\ref{eq3a}, \ref{eq4a}) result from (\ref{MATEL}) directly.
Eqs.~(\ref{eq1a},\ref{eq2a}) are known for a long time,
cf.~ref.~\cite{CG}.
\subsection{Vector Leptoquarks}
\label{sect2C}
While for the case of vanishing anomalous couplings,
$\kappa_G = \lambda_G \equiv 0$, the pair production cross section
follows from eq.~(\ref{MATEL})  it cannot be
derived by the technique discussed in section~\ref{sectXX}
for finite anomalous couplings because
eq.~(\ref{eqdelt}) does not hold. For the
general case we have performed the calculation
of $|{\cal M}^{V}_{q,g}|^2$ by
{\sf CompHEP}~\cite{COMPHEP} in the Feynman gauge\footnote{The
Feynman rules given in appendix A have been implemented into
{\sf CompHEP} as a new model.}.
We also checked gauge invariance explicitly
using {\tt FORM}~\cite{FORM} in the $R_{\xi}$~gauge
with a free gauge parameter. For  these calculations,
the diagrams in figure~1 must be supplemented
by an initial state ghost contribution.

The differential and integral pair production cross sections for
$gg$ scattering are
\begin{equation}
\label{eq3b}
\frac{d \hat{\sigma}_{V\overline{V}}^{gg}}{d \cos \theta}
= \frac{\pi \alpha_s^2}{192 \sh} \beta \sum_{i=0}^{14}  \chi_i^g(\KG,\LG)
\frac{F_i(\sh,\beta,\cos \theta)}{(1 - \beta^2 \cos^2 \theta)^2},
\end{equation}
with
\begin{equation}
\begin{array}{lcrcrcrcrcr}
{\displaystyle{
\sum_{i=0}^{14}}} \chi_i^g(\KG,\LG) F_i
&=& F_0       &+& \KG F_1    &+& \LG F_2
&+& \KG^2 F_3     &+& \KG \LG F_4  \\

\vspace*{3mm}
\noindent
&+& \LG^2 F_5 &+& \KG^3 F_6  &+& \KG^2 \LG F_7
&+& \KG \LG^2 F_8 &+& \LG^3 F_9 \\

\vspace*{3mm}
\noindent
&+& \KG^4 F_{10} &+& \KG^3 \LG F_{11}
&+& \KG^2 \LG^2 F_{12}  &+& \KG \LG^3  F_{13} &+&  \LG^4 F_{14},
\end{array}
\end{equation}
\begin{equation}
\label{eq4b}
\hat{\sigma}_{V\overline{V}}^{gg}
= \frac{\pi \alpha_s^2}{96 M_V^2}  \sum_{i=0}^{14}  \chi_i^g(\KG,\LG)
\widetilde{F}_i(\sh,\beta),
\end{equation}
\begin{equation}
\widetilde{F}_i = \frac{M_V^2}{\sh}
\int_0^{\beta} d \xi \frac{F_i(\xi = \beta \cos \theta)}
{(1 - \xi^2)^2}.
\end{equation}
The functions $F_i(\sh,\beta,\cos \theta)$ and
$\widetilde{F}_i(\sh,\beta)$ are obtained after a lengthy calculation
and are given in appendix~B.
Similar
to the case of $\gamma g$~\cite{BBP} and $\gamma \gamma$~\cite{BB94}
scattering, the contributions {\it linear} in either $\kappa_G$ or
$\lambda_G$ do not contain unitarity violating pieces
$\propto \hat{s}/M_V^2$.

For $q\overline{q}$  scattering the cross section reads
\begin{equation}
\label{eq1b}
\frac{d \hat{\sigma}_{V\overline{V}}^{q\overline{q}}}{d \cos \theta}
= \frac{2 \pi \alpha_s^2}{9 M_V^2} \beta^3
\sum_{i=0}^{5}  \chi_i^q(\KG,\LG)
G_i(\sh,\beta,\cos \theta),
\end{equation}
with
\begin{equation}
\sum_{i=0}^{5}  \chi_i^q(\KG,\LG) G_i  =  G_0  +  \KG G_1  +  \LG G_2
                                       +  \KG^2 G_3  +  \KG \LG G_4
 +  \LG^2 G_5.
\end{equation}
The integrated cross section is
\begin{equation}
\label{eq2b}
\hat{\sigma}_{V\overline{V}}^{q\overline{q}}
= \frac{4 \pi \alpha_s^2}{9 M_V^2} \B^3
\sum_{i=0}^5 \chi_i^q(\KG,\LG)
\widetilde{G}_i(\sh,\beta),
\end{equation}
where
\begin{equation}
\widetilde{G}_i =
\int_0^1 d \cos \theta~~G_i(\sh, \B, \cos \theta).
\end{equation}
The functions $G_i(\sh,\beta,\cos \theta)$
and $\widetilde{G}_i(\sh,\beta)$ are
listed in appendix~B.  For $\LG = 0$
eqs.~(\ref{eq1b},\ref{eq2b}) agree
with those found in \cite{SCAL4}
specifying the color factor\footnote{Note  that this result disagrees
with eq.~(4) in \cite{EBO94} by a factor of $81/4$.}.
Relations for
the special cases $\KG = 0$
\cite{ABB} and $\KG = 1$ \cite{BR} have been obtained for other reactions
previously.

Contrary to the case of gluo-production~(\ref{eq4b}), terms $\propto
\hat{s}/M_V^2$ are contained even in the contribution $G_0$
because
we did not impose a relation between the fermionic and bosonic
couplings of the leptoquarks for Yang-Mills type leptoquark-gauge boson
 couplings. To
restore unitarity,  graphs with lepton exchange
would have to be added
for the process $q\overline{q} \rightarrow \Phi_V\overline{\Phi}_V$.
The value required for the adjusted fermionic couplings, however,
would be too large to be consistent
with the limits derived in~\cite{LEUR}.  The Lagrangian~(2) is assumed
to parametrize leptoquark interactions for not too
large energies, i.e. in the threshold range.
It has to be supplemented by further terms
restoring the correct high energy behaviour for
$\hat{s} \gg M^2_V~~$. These
terms are model dependent and are related
to the specific scenario leading to leptoquarks in the mass range
of $100 \GeV$ to $\sim 1 \TeV$. Approaching high energies,
symmetry breaking
scales are passed and the respective
Higgs terms contribute.
\section{Production Cross Sections}
\label{sect1B}
Subsequently we will calculate the
differential and integral hadronic
production cross sections for leptoquark pair production at different
colliders. In the present paper
we will apply the collinear
parton model to describe the initial
state of the hard scattering process $f_i f_j \rightarrow
\Phi \overline{\Phi}$. The densities $f_i$ are the
parton densities in the
case of the $p\overline{p}$ or
$pp$ collisions. For the resolved photon contributions in
$ep$ collisions one of the densities
is the probability for finding a quark, antiquark, or gluon in the
electron in a neutral current process. These distributions are
described by the convolution
\begin{equation}
f_{i/e}(x,\mu^2) = [f_{\gamma/e}(\mu^2) \otimes f_{i/\gamma}(\mu^2)](x),
\end{equation}
where
$f_{\gamma/e}(x,\mu^2)$ is the photon density in an electron
and
$f_{i/\gamma}(x,\mu^2)$ denotes  the
density of parton $i$ in the photon. The
convolution of the densities is given by
\begin{equation}
[A \otimes B](x) = \int_0^1 \int_0^1 dx_1 dx_2 \delta(x - x_1 x_2)
A(x_1) B(x_2).
\end{equation}
In the same way the parton densities for photoproduction of leptoquark
pairs at $e^+e^-$ colliders are described. For $\gamma \gamma$ colliders,
at which the photon beams are prepared by laser back--scattering, the
distribution $f_{\gamma/e}(z)$ is given by the Compton spectrum
$\phi_C(z)$, eq.~(\ref{fCOMP}).

In  $ep$,
$\gamma^* \gamma^*$,  and $\gamma \gamma$
scattering, in addition to
the
resolved photon
subprocesses, {\it direct} contributions due to $\gamma g$
and $\gamma \gamma$ fusion are present which have been studied in
refs.~\cite{BBP} and
\cite{JB93,BB94}, respectively.

In terms of the  {\it generalized} partonic distributions
$f_i^{(c)}(x_c, \mu^2)$
the differential cross section reads:
\begin{equation}
\label{eq30}
\frac{d^2 \sigma}{d \eta d p_{\perp}^2} = \int_0^1 dx_a
\frac{\sqrt{S}~\theta(\sh - 4 \MP)}
{x_a \sqrt{S} - \sqrt{\MP + p_{\perp}^2} (\ch \eta + \shh \eta)}
x_a x_b f_i^{(a)}(x_a, \mu^2) f_j^{(b)}(x_b, \mu^2)
\frac{2}{\sh \beta} \frac{d \hat{\sigma}^{ij}(\sh,\thh,\uh)}
{d \cos \theta}.
\end{equation}
Here, $\eta$ denotes the rapidity of one of the leptoquarks,
\begin{equation}
\eta = \frac{1}{2}  \ln \left | \frac{E_h + p_{zh}}{E_h - p_{zh}}
\right |,
\end{equation}
with $E_h$ and $p_{zh}$ the leptoquark energy  and longitudinal momentum.
$p_{\perp}$ is the
 transverse momentum in the laboratory frame,
 $\sh = x_a x_b s$, and
\begin{equation}
\label{eq31}
x_b = \frac{x_a
\sqrt{\MP + p_{\perp}^2}~(\ch \eta - \shh \eta)}{x_a \sqrt{s}
- \sqrt{\MP + p_{\perp}^2}~(\ch \eta + \shh \eta)}.
\end{equation}
The Mandelstam variables $\thh$ and $\uh$ in the cms are
\begin{eqnarray}
\thh &=& \MP - x_a \sqrt{s} \sqrt{M^2 + p_{\perp}^2}
(\ch \eta - \shh \eta)
= \MP - \frac{\sh}{2} (1 - \beta \cos \theta),
\\
\uh &=& \MP - x_b \sqrt{s} \sqrt{\MP + p_{\perp}^2}
(\ch \eta + \shh \eta)
= \MP - \frac{\sh}{2} (1 + \beta \cos \theta).
\end{eqnarray}
The differential cross sections in the partonic sub-systems
$d \hat{\sigma}^{ij}/d \cos \theta$
have been given in eqs.~(\ref{eq3a},\ref{eq1a},\ref{eq3b},\ref{eq1b})
in section~\ref{PARCS}.
The single differential distributions $d \sigma/d\eta$ and
$d \sigma/d p_{\perp}^2$ are derived from eq.~(\ref{eq30}).
Eq.~(\ref{eq31})
constrains the rapidity and $p_{\perp}$ ranges to
\begin{eqnarray}
- \frac{1}{2} \log \frac{1+B}{1-B} \leq \eta \leq
  \frac{1}{2} \log \frac{1+B}{1-B}, \\
0 \leq p_{\perp} \leq \frac{\sqrt{s}}{2} \sqrt{
\beta^2  -
\thhh^2 \eta
},
\end{eqnarray}
where $B = \sqrt{1 - 4(\MP + p_{\perp}^2)/s}$.

Finally
the integral cross sections are
\begin{equation}
\sigma(\Phi \overline{\Phi}) = \int_0^1 \int_0^1 dx_a dx_b
 f_i^{(a)}(x_a, \mu^2) f_j^{(b)}(x_b, \mu^2)
\hat{\sigma}^{ij}(\sh)~\theta(\sh - 4 \MP).
\end{equation}
To be specific, we list the relations
for the different contributions to the
integrated
cross sections for $pp  (p\overline{p})$,
$ep$ and $\gamma \gamma (\gamma^* \gamma^*)$ scattering
explicitly below.

\subsection{$p\overline{p}$ and $pp$ scattering}

\vspace{1mm}
\noindent
Here the total cross section consists  of
contributions from
quark--antiquark annihilation and gluon--gluon fusion
\begin{equation}
\sigma_{S,V}^{pp}(s,M_{\Phi}) = \sigma_{S,V}^{pp;q}(s,M_{\Phi})
+\sigma_{S,V}^{pp;g}(s,M_{\Phi}),
\end{equation}
where
\begin{eqnarray}
\sigma_{S,V}^{pp;q}(s,M_{\Phi}) &=&
\sum_{f=1}^{N_f} \int_0^1 dx_1 \int_0^1 dx_2
\left [ q_f(x_1,\mu) \overline{q}_f(x_2,\mu)
+
 \overline{q}_f(x_1,\mu) q_f(x_2,\mu) \right ] \nonumber\\
&\times &
\hat{\sigma}_{S,V}^{q}(\sh,M_{\Phi})~\theta(\hat{s} - 4 M_{\Phi}^2),
\\
\sigma_{S,V}^{pp,g}(s,M_{\Phi}) &=&
\int_{0}^1 dx_1 \int_0^1 dx_2
~G(x_1,\mu)G(x_2,\mu)
\hat{\sigma}_{S,V}^{g}(\sh,M_{\Phi})~\theta(\hat{s} - 4 M_{\Phi}^2).
\end{eqnarray}
$q_f(x,\mu), \overline{q}_f(x,\mu)$, and $G(x,\mu)$ denote the
quark, antiquark and gluon
distributions of the proton (antiproton),
and  $\mu$ is the factorization scale.

\subsection{$ep$ scattering}

\vspace{1mm}
\noindent
In  $ep$ scattering, the two contributions
to the
production cross section are
the direct process $\gamma^* g \rightarrow
\Phi\overline{\Phi}$~\cite{BBP} and the
resolved photon process. Due to
the photon-leptoquark coupling, the direct contribution, $\sigma_{dir}$,
contains a factor
$Q_{\Phi}^2$ while the resolved one, $\sigma_{res}$,
does not depend on the leptoquark charge.
We use the Weizs\"acker--Williams approximation (WWA) to describe the
photon spectrum both in the case of the direct and resolved photon
contributions.
This approximation is known to hold at an accuracy of 10 to 15~\%.

The total cross section is
\begin{equation}
\sigma_{S,V}^{ep,tot} = \sigma_{S,V}^{ep,dir} + \sigma_{S,V}^{ep,res},
\end{equation}
with (cf.~\cite{BBP})
\begin{equation}
\sigma_{S,V}^{ep,dir} = \int_{y_{min}}^{y_{max}} dy
                        \int_{x_{min}}^{x_{max}} dx \phi_{\gamma/e}(y)
                        G_{p}(x,\mu^2)
\hat{\sigma}_{S,V}^{dir}(\hat{s},M_{\Phi})
\theta(\hat{s} - 4M^2_{\Phi}),
\end{equation}
where
$x_{min} = 4M^2_{\Phi}/yS$, $\hat{s} = S x y$,
$S=4 E_e E_p$,
$x_{max} = 1$,
$y = P.q/ P.l_e$, with $q = l_e - l'_e$, and
$P, l_e, l'_e$ the four momenta of the
proton, the incoming and outgoing electron.
The boundaries $y_{min,max}$ are given by
\begin{equation}
\label{ymM}
y_{min,max} = \frac{S + \widetilde{W}^2 \pm \sqrt{(S - \widetilde{W}^2)^2
- 4m_e^2 \widetilde{W}^2}}{2(S + m_e^2)},
\end{equation}
%
where  $\widetilde{W}^2 = (2 M_{\Phi} + m_p)^2 - m_p^2$,   and
$m_e$ and $m_p$ are the electron and proton mass, respectively.
$\phi_{\gamma/e}(y)$ denotes the Weizs\"acker-Williams
distribution, see e.g.~\cite{FRIX}~:
%
\begin{equation}
\label{eqphi}
\phi_{WWA}(y) = \frac{\alpha}{2 \pi} \left [ 2 m_e^2 y \left (
\frac{1}{Q^2_{max}} -
\frac{1}{Q^2_{min}} \right ) + \frac{1 + (1 - y)^2}
{y} \log \frac{Q^2_{max}}{Q^2_{min}} \right ].
\end{equation}
%
To parametrize the scales $Q_{min,max}^2$ we choose the kinematic
limits
\begin{equation}
Q^2_{min} = \frac{m_e^2 y^2}{1 - y},~~~~~~~~~~~~~~~Q^2_{max} = y S
- 4 \MP
- 4 M_{\Phi} m_p,
\end{equation}

The cross sections in the $\gamma^* g$
subsystem are
\begin{equation}
\hat{\sigma}_{S,V}^{dir}(\hat{s},M_{\Phi})
= \frac{\pi \alpha \alpha_s(\mu^2)}
{\hat{s}} Q^2_{\Phi} R_{S,V}(\hat{s},M_{\Phi}),
\end{equation}
where
\begin{eqnarray}
\label{eqRS}
R_S &=& (2 - \beta^2)\beta - \frac{1}{2}(1 - \beta^4)\XL, \\
\label{eqRV}
R_V &=& \sum_{j=0}^{20}
\chi_j^*(\kappa_{\gamma},\kappa_G,\lambda_{\gamma},\lambda_G)
\widetilde{H}_j(\hat{s},\beta).
\end{eqnarray}
The functions
$\chi_j^*(\kappa_{\gamma},\kappa_{G},\lambda_{\gamma},\lambda_{G})$
and
$\widetilde{H}_j$ were given in ref.~\cite{BBP} in~eqs.~(12;~A.2) where
we defined
$\widetilde{H}_j \equiv (M_V^2/\hat{s}) \widetilde{F}_j^*$~\footnote{
We have labeled the functions  $\chi_j$ and
$\widetilde{F}_j$ of \cite{BBP} (A.2) by a star to distinguish them from
those given in appendix~B.
The expression for
$\widetilde{F}^{18}$ in~\cite{BBP}
contains a typographical error, the factor $(1 - 6 \beta^2)$ there
should read
$(1 + 6 \beta^2)$.}.

The cross sections due to the resolved photon contributions read:
%
\begin{eqnarray}
\label{xsep}
\sigma_{S,V}^{ep, res}(s,M_{\Phi}) &=&
\int_{y_{min}}^{y_{max}} dy
\int_{4 \MP/Sy}^1 dz
\int_{4 \MP/Syz}^1 dx
\phi_{\gamma/e}(y)
\theta(\hat{s} - 4 M_{\Phi}^2)
\nonumber\\
&\times&
\Biggl \{
\sum_{f=1}^{N_f} \left [ q_f^{\gamma}(z, \mu_1)
\overline{q}_f^p(x,\mu_2) +
 \overline{q}_f^{\gamma}(z, \mu_1)
q_f^p(x,\mu_2) \right ]
                 \hat{\sigma}_{S,V}^q(\sh, M_{\Phi})
\nonumber\\
&+&
G^{\gamma}(z, \mu_1) G^p(x,\mu_2)
                 \hat{\sigma}_{S,V}^g(\sh, M_{\Phi})
 \Biggr \}.
\end{eqnarray}
%
Here,
$q^{\gamma,p}$ and $G^{\gamma,p}$ denote the quark and gluon densities
in the photon and proton, respectively, and
$\sh = x y z S$. $\mu_1$ and $\mu_2$ are  the mass
factorization scales of the photon and  proton distributions,
which are different in general.
The cross sections for the hadronic subprocesses
$\hat{\sigma}^{q,g}_{S,V}$
are given
in eqs.~(\ref{eq4a},\ref{eq2a},\ref{eq4b},\ref{eq2b}).

\subsection{ $\gamma \gamma$  fusion}
\noindent
For $\gamma \gamma$ or $\gamma^* \gamma^*$
scattering three terms contribute to the cross
section: the direct process
$\gamma \gamma \rightarrow \Phi\overline{\Phi}$, $\sigma_{dir}$,
a term in which
one of the photons is resolved and the second couples directly to
the leptoquarks, $\sigma_{dir/res}$, and the double resolved
contribution, $\sigma_{res}$,
\begin{equation}
\sigma_{S,V}^{\gamma \gamma, tot} =
\sigma_{S,V}^{\gamma \gamma, dir} +
\sigma_{S,V}^{\gamma \gamma, dir/res} +
\sigma_{S,V}^{\gamma \gamma, res}.
\end{equation}
The third term~(eq. (\ref{xsgg}))
is charge independent, but the first and the second terms behave
$\propto Q_{\Phi}^4$ and
$\propto Q_{\Phi}^2$, respectively.
The cross section for the direct contribution reads~\cite{BB94}
\begin{equation}
\label{xsggdir}
\sigma_{S,V}^{\gamma \gamma, dir}
(s,M_{\Phi}) = \int_{y_{min}/y_{max}}^{y_{max}} dy_1
                 \int_{y_{min}/y_1}^{y_{max}} dy_2
\Phi_{\gamma/e}(y_1)
\Phi_{\gamma/e}(y_2)
                 \hat{\sigma}_{S,V}^{dir}(\sh, M_{\Phi})
\theta(\hat{s} - 4 M_{\Phi}^2).
\end{equation}
Here the subsystem cross sections are:
\begin{equation}
\hat{\sigma}_{S,V}^{dir}(\hat{s},M_{\Phi})  =  \frac{\pi \alpha^2}
{\hat{s}} Q^4_{\Phi} N_c R_{S,V}^*(\hat{s},M_{\Phi}),
\end{equation}
with $\hat{s} = y_1 y_2 S, S = 4 E_{e^+} E_{e^-}$, $N_c = $,
and
\begin{eqnarray}
\label{eq55a}
R^*_S &=& 2 R_S, \nonumber\\
R^*_V &=& 2 \sum_{j=0}^{20}
\chi_j^*
(\kappa_{\gamma},\kappa_{\gamma},\lambda_{\gamma},\lambda_{\gamma})
\widetilde{H}_j(\hat{s},\beta).
\end{eqnarray}
Effectively $R_V^*$ depends on only 15 independent functions~\cite{BB94}
due to the symmetry of the function
$\chi_j^*
(\kappa_{\gamma},\kappa_{\gamma},\lambda_{\gamma},\lambda_{\gamma})$
in eq.~(\ref{eqRV}).

The direct--resolved term is given by
\begin{eqnarray}
\label{eq55b}
\sigma_{S,V}^{\gamma \gamma, dir/res}
(s,M_{\Phi}) &=& 2 \int_{y_{min}/y_{max}}^{y_{max}} dy_1
                 \int_{y_{min}/y_1}^{y_{max}} dy_2
                 \int_{4 \MP/S y_1 y_2}^1 dz
\Phi_{\gamma/e}(y_1)
\Phi_{\gamma/e}(y_2) G_{\gamma}(z,\mu)
 \nonumber\\
&\times&
                 \hat{\sigma}_{S,V}^{\gamma g}(\sh, M_{\Phi})
\theta(\hat{s} - 4 M_{\Phi}^2).
\end{eqnarray}
with $\mu$ the factorization scale.
Because of the small size of the couplings $\lambda_{lq} \ll e$
only the subprocess due to gluon-photon fusion contributes.

Finally
the doubly--resolved contribution reads:
\begin{eqnarray}
\label{xsgg}
\sigma_{S,V}^{\gamma \gamma, res}
(s,M_{\Phi}) &=& \int_{y_{min}/y_{max}}^{y_{max}} dy_1
                 \int_{y_{min}/y_1}^{y_{max}} dy_2
                 \int_{4 \MP/S y_1 y_2}^1 dz_1
                 \int_{4 \MP/S y_1 y_2 z_1}^1 dz_2
\Phi_{\gamma/e}(y_1)
\Phi_{\gamma/e}(y_2) \nonumber\\
&\times&
\Biggl \{
\sum_{f=1}^{N_f} \left [ q_f^{\gamma}(z_1, \mu_1)
\overline{q}_f^{\gamma}(z_2, \mu_2) +
 \overline{q}_f^{\gamma}(z_1, \mu_1)
q_f^{\gamma}(z_2, \mu_2) \right ]
                 \hat{\sigma}_{S,V}^q(\sh, M_{\Phi})
 \nonumber\\
&+&
G^{\gamma}(z_1, \mu_1) G^{\gamma}(z_2, \mu_2)
                 \hat{\sigma}_{S,V}^g(\sh, M_{\Phi})
 \Biggr \}
\theta(\hat{s} - 4 M_{\Phi}^2).
\end{eqnarray}
%
where $\sh = y_1 y_2 z_1 z_2 S$.

In
$e^+ e^-$ scattering the
functions $\Phi_{\gamma/e}(y_i)$
denote the Weizs\"acker--Williams distribution (\ref{eqphi}) with
the parameters
\begin{equation}
Q^2_{i,min} = \frac{m_e^2 y_i^2}{1 - y_i},~~~~~~~~Q^2_{i,max} =
S y_i - 4 M^2_{\Phi} - 4 M_{\Phi} m_e,
\end{equation}
\begin{equation}
\label{eq46}
y_{min,max} = \frac{S + \widetilde{W}^{*2} \pm \sqrt{(S -
\widetilde{W}^{*2})^2
- 4m_e^2 \widetilde{W}^{*2}}}{2(S + m_e^2)},
\end{equation}
and $W^{*2} = (m_e + 2 M_{\Phi})^2 - m_e^2$.

For a $\gamma \gamma$ collider operating with photon beams which are
produced by laser back scattering, the spectrum $\Phi_{\gamma/e}(y)$ is
given by
\begin{equation}
\phi_C(y) = \frac{1}{N(x)}
\left [ 1 - y  + \frac{1}{1 - y} - \frac{4 y}{x(1 - y)}
+ \frac{4y^2}{x^2 (1-y)^2} \right ],
\label{fCOMP}
\end{equation}
with
\begin{equation}
N(x) = \frac{16+32x + 18x^2 + x^3}{2x (1+x)^2} + \frac{x^2 - 4x -8}
{x^2} \ln(1 + x),
\end{equation}
and
\begin{equation}
\label{eq48a}
0 = y_{min} \leq y \leq y_{max} = x/(x+1),~~~~x = 2(\sqrt{2} +1),
\end{equation}
cf.~\cite{NOVO}.
Here complete beam conversion is assumed for the laser back scattering
process. We use eq.~(\ref{fCOMP})
as an approximate description, which may be
needed to be refined according to different
technical aspects at future $\gamma \gamma$ colliders~\cite{GIN}.

%
\section{Numerical Results}
\label{sect11}
\subsection{$p\overline{p}$ and $pp$  scattering}

\vspace{1mm}
\noindent
The different numerical values of the cross sections for leptoquark pair
production at $p\overline{p}$ and $pp$ colliders  calculated
below are determined by the different resulting parton--parton
luminosities $\tau d {\cal L}/ d \tau$.
In the collinear picture described above,
the differential luminosities are
\begin{eqnarray}
\tau \frac{d {\cal L}^q}{ d \tau} &=& \tau \int_{\tau}^1 \frac{dx}{x}
\sum_{i=1}^{N_f}
\left [
f_{q_i/a}(x, \mu_1) f_{\overline{q}_i/b}(\frac{\tau}{x}, \mu_2)
+ f_{\overline{q}_i/a}(x, \mu_1) f_{q_i/b}(\frac{\tau}{x}, \mu_2)
\right ],  \\
\tau \frac{d {\cal L}^g}{ d \tau} &=& \tau \int_{\tau}^1 \frac{dx}{x}
f_{g/a}(x, \mu_1) f_{g/b}(\frac{\tau}{x}, \mu_2),
\end{eqnarray}
with  $\tau = \sh/s$, and $\mu_1$ and $\mu_2$ are the corresponding
factorization scales.
In figure~3  a comparison of the quark-antiquark and gluon-gluon
luminosities is given for the kinematic   range at the
TEVATRON and LHC.
We used
the  distributions~\cite{CTEQ3}\footnote{Other parametrizations of the
parton densities in the proton~\cite{GRVMRS95} agree very well
with the  parametrization~\cite{CTEQ3}.}
to describe the
parton densities of the proton and choose
$\mu_1 = \mu_2 = \sqrt{\sh} \equiv \sqrt{\tau s}$ in the following.

At a fixed value of $\tau$
the differential quark-antiquark luminosity
$\tau (d {\cal L}^q/d\tau)$ at the
TEVATRON
is much larger than  at LHC. The corresponding values for
$\tau (d {\cal L}^g/d\tau)$ are
rather similar. Since the dominant contributions
to the scattering cross sections at LHC are due to lower $\tau$ values
when
compared to the range at the
TEVATRON, large contributions to the
scattering cross section are  expected for the
gluon fusion process.
On the other hand, the quark-antiquark annihilation process is expected
to yield a large contribution to the scattering cross section at the
TEVATRON.

In figures~4a,b, the pair production cross sections for scalar leptoquarks
at the
TEVATRON and LHC are shown for the mass range $M_S \geq 100 \GeV$.
The respective contributions to the cross section
due to quark-antiquark and gluon-gluon scattering are basically
a consequence of the behaviour of the differential parton luminosities:
whereas the quark terms yield the largest contributions to the
cross section at the
TEVATRON in
this mass range, at LHC the dominant contribution is due to gluon-gluon
fusion. The quark contributions become important also at LHC energies
for large masses.

The production cross sections for vector leptoquarks at the
TEVATRON and
LHC are shown in figures~5a,b. We observe a sizeable dependence of
the cross section on the value of both
the anomalous couplings $\kappa_G$
and $\lambda_G$. Moreover, their impact on the quark and gluon
contributions turns out to be different.
As in the case of scalar leptoquarks
the production cross section at the
TEVATRON, figure~5a,
is dominated by the quark
contributions. On the other hand the gluonic terms dominate in the
case of LHC, figure~5b.
The numerical values of the
cross sections agree with those calculated
in~\cite{VECT1}~(fig.~18,19),~\cite{VECT2}~\footnote{A recent numerical
update was given in ref.~\cite{UPDATE}.}, for selected values of
$\kappa_G = 1,0$ and $\lambda_G \equiv 0$ treated there,
within the uncertainty due to the different parton densities used.

We  vary both $\kappa_G$ and $\lambda_G$ to illustrate
the overall behaviour and consider all combinations $\kappa_G,
\lambda_G~$ at a fixed, characteristic leptoquark mass.
These variations lead to changes of about two orders  of magnitude
in the production cross sections both at LHC and the
TEVATRON allowing
for
$-0.5 \leq \kappa_G < 3.5$ and $\lambda_G < 1$.
In figure~6a,b the dependence of the total production cross section
is shown for the case of the
TEVATRON and LHC, assuming $M_{LQ} = 150 \GeV$
and $M_{LQ} = 500 \GeV$, respectively.
In neither case is
the minimal cross section  obtained for anomalous
couplings close to the Yang-Mills type couplings. For
TEVATRON energies the minimal cross section is
obtained for $\lambda_G = -0.208, \kappa_G = 1.3$
choosing $M_{V} = 150 \GeV$ as an example.
On the other hand, at LHC for $M_{V} = 500 \GeV$
the minimal cross section is obtained for $\lambda_G = -0.052,
\kappa_G = 1.02$. The latter  values are rather close to those in the
case of a minimal vector coupling.

For the determination of the leptoquark signal both the rapidity and
$p_{\perp}$ distributions may be used. They are shown in figures~7 and
8 for the kinematic   ranges at the
TEVATRON and LHC assuming the above
leptoquark masses as examples.
The rapidity distributions for both scalar and vector leptoquark
pair production are wider at LHC
compared to those at the
TEVATRON,
which are more peaked at $|\eta| \sim 0$. The $p_{\perp}$ distribution
for scalar
and vector leptoquarks at the
TEVATRON (figures 8~a,b) peak at
$p_{\perp} \sim 100 \GeV$, while the corresponding
$p_{\perp}$ distributions at LHC are much wider and peak positions of
$p_{\perp} \sim 150 \GeV$ for scalar  and
$p_{\perp} \sim 300 \GeV$ for vector leptoquarks with a
minimal vector coupling are found.

The estimates given
are  illustrative and can not replace
a complete analysis which accounts for
specific experimental and detector details in the respective experiments
and a detailed investigation of background reactions.
This  is not the intention of the present paper.
The numbers given below for accessible mass ranges~(section~5.4)
are therefore
meant to be indicative since they are based on the signal events only.
They  serve as an illustration of the principal search
potential  in different reactions and at different colliders.

\subsection{$ep$  scattering}

\vspace{1mm}
\noindent
Figure~9a shows the dependence of the integrated production cross section
on the leptoquark mass for scalar pair production at HERA for both
$|Q_{\Phi}| = 5/3$ and $1/3$. For the calculation of the resolved photon
contribution we used the parametrization~\cite{GRVPH} to describe the
parton densities of the photon. The mass dependence of the cross sections
in the range $M_{\Phi} > 40 \GeV$ is found to be nearly exponential.
For low charge
leptoquarks the resolved photon terms dominate, whereas for leptoquarks
with a high charge the largest contribution to the cross section is due
to the direct term. A similar behaviour is observed in a more extended
mass range for the same process at
LEP~$\otimes$~LHC, see figure~9b.

For the case of vector leptoquark pair production we consider different
choices of
the anomalous couplings in the range
$\kappa,~\lambda~\epsilon~[-1,+1]$, and put $\kappa_{\gamma} = \kappa_G$
and $\lambda_{\gamma} = \lambda_G$ always.

For the kinematic   range of HERA figure~10a shows the cross sections
for vector leptoquark pair production for a selection of anomalous
couplings.
Rather large cross sections are obtained for the
choice $\kappa_{\gamma,G} = \lambda_{\gamma,G} = -1$.
The pair production
cross sections are larger in the case of Yang-Mills type
gauge boson-leptoquark couplings $\kappa_{\gamma,G} =
\lambda_{\gamma,G} = 0$ than in the case of minimal couplings
$\kappa_{\gamma,G} = 1, \lambda_{\gamma,G} = 0$.
Although the accessible mass range at HERA is smaller than that at
the
TEVATRON, a search for vector leptoquark pair production at HERA for
third
generation leptoquarks appears to be worthwhile, cf.~\cite{PDG}.
In  $ep$
scattering, besides
 the anomalous couplings of the gluon also
those of the photon are probed
and corresponding constraints can be derived.

An analogous behaviour of the scattering cross sections, covering
a wider mass range, is obtained for the same processes at
LEP~$\otimes$~LHC, see figure~10b.

\subsection{$\gamma \gamma$  scattering}

\vspace{1mm}
\noindent
The different contributions to the integrated pair production cross
section for scalar leptoquarks at a future $e^+e^-$ linear collider
operating at $\sqrt{s} = 500 \GeV$ are depicted in figure~11a. For
leptoquarks carrying a charge $|Q_{\Phi}| = 5/3$ the direct process
in the reaction $e^+e^- \rightarrow \Phi \overline{\Phi} X$~~\footnote{
Here we do not consider the contributions of the
process $e^+e^- \rightarrow \Phi \overline{\Phi}$ which was
dealt with in ref.~\cite{BR} previously. For
a detailed investigation of this process for
the case of vector leptoquarks with general
anomalous couplings $\kappa_{\gamma,Z}, \lambda_{\gamma,Z}$, see
ref.~\cite{BBK96A}.}
dominates at lower values of $\beta$. Here the virtual photon spectra
were approximated by the Weizs\"acker--Williams
distribution,~eq.~(\ref{eqphi}).
At large $\beta$ the cross section
is further enhanced by the resolved photon
contributions. On the other hand,
for low charge leptoquarks as $|Q_{\Phi}| = 1/3$, the direct contribution
is suppressed by orders of magnitude against the resolved photon terms.

In figure~11b the contributions to the production cross sections are
shown for scalar leptoquarks at a $\gamma \gamma$ collider at
$\sqrt{s} = 500 \GeV$. The general observation is the same as in
the previous case but at lower values of $\beta$ larger cross sections
are obtained. For leptoquarks of a charge $|Q_{\Phi}| = 5/3$ thus nearly
the complete accessible phase space can be probed.

As in the case of $ep$ scattering we consider a series of anomalous
couplings for  vector leptoquark pair production. Numerical
results are illustrated for the process
$\gamma^* \gamma^* \rightarrow V \overline{V}$ at an $e^+e^-$
linear collider at $\sqrt{s} = 500 \GeV$ assuming Weizs\"acker-Williams
spectra for the virtual photons, in figure~12a.
Among the different choices of anomalous couplings considered,
the smallest cross sections were obtained for the minimal vector
couplings. The corresponding cases for a $\gamma \gamma$ collider
operating at the same cms energy are depicted in figure~12b. For all
three choices of the anomalous couplings, the accessible search range
reaches nearly the kinematic boundary for $|Q_{\Phi}| = 5/3$
leptoquarks, while for
$|Q_{\Phi}| = 1/3$ leptoquarks the corresponding limits are smaller.

\subsection{A comparison of the search potential at different colliders}

\vspace{1mm}
\noindent
The leptoquark search limits  which can be reached for the reactions
discussed in the previous sections at different high energy colliders
are summarized in
table~2. Estimates of the accessible mass ranges are given based
on 10 and 100 signal events, respectively.
Here we consider the mass range above $45 \GeV$, which was already
widely excluded by the searches at LEP~1, cf.~\cite{LQLEP}.

For hadron colliders such as the
TEVATRON and LHC the search limits are
flavor and family independent. Since in the other reactions
also
photon-leptoquark interactions contribute\footnote{Additional
contributions due to Higgs-, $Z$- and $W^{\pm}$-boson couplings
are not considered in the present paper.}, the respective event rates
depend partly on the leptoquark charge.

\vspace{7mm}
\noindent
\begin{center}
\begin{tabular}{||l|c|r|r||c|c|c||c|c||}
\hline  \hline
\multicolumn{1}{||c|}{          } &
\multicolumn{1}{c|  }{          } &
\multicolumn{1}{c|  }{          } &
\multicolumn{1}{c|| }{          } &
\multicolumn{1}{c| } {          } &
\multicolumn{2}{c||}{Scalar            } &
\multicolumn{2}{c||}{Vector            } \\
\multicolumn{1}{||c|}{Collider  } &
\multicolumn{1}{c|  }{Mode      } &
\multicolumn{1}{c|  }{$\sqrt{S}$} &
\multicolumn{1}{c|| }{Luminosity} &
\multicolumn{1}{c| } {Q         } &
\multicolumn{2}{c||}{       Leptoquarks} &
\multicolumn{2}{c||}{       Leptoquarks} \\
\cline{6-9}
& & & & &
\multicolumn{1}{c|  }{$ 100 \#$ } &
\multicolumn{1}{c|| }{$ 10 \#$  } &
\multicolumn{1}{c|  }{$ 100 \#$ } &
\multicolumn{1}{c|| }{$ 10 \#$  } \\
\hline  \hline
TEVATRON & $p\overline{p}$ & 1.8 TeV & $100 pb^{-1}$ & & 140& 200& 170
& 225\\
\hline
TEV33 & $p\overline{p}$ & 2.0 TeV & $1 fb^{-1}$ & & 210& 290& 290& 370\\
\hline
LHC    & $pp$ & 14 TeV & $10 fb^{-1}$ &      & 900 & 1200& 1200& 1500\\
\hline
HERA & $ep$ & 314 GeV & $100 pb^{-1}$ & $1/3$&  -   & 50   &  50  & 60  \\
\cline{5-9}
         &      &         &          & $5/3$ & 45  & 60  & 60  & 75  \\
\cline{4-9}
     &      &         & $500 pb^{-1}$ & $1/3$& 45   & 60   &  60  & 75  \\
\cline{5-9}
         &      &         &          & $5/3$ & 55  & 75  & 70  & 85  \\
\hline
LEP~$\otimes$~LHC
     & $ep$ & 1.26 TeV & $1 fb^{-1}$ & $1/3$ & 125 & 180 & 180 & 240 \\
\cline{5-9}
         &      &         &          & $5/3$ & 165 & 225 & 210 & 270 \\
\hline
LINAC & $\gamma^* \gamma^*$
    & 500  GeV & $10 fb^{-1}$        & $1/3$ & 90  & 120 & 120 & 155 \\
\cline{5-9}
$e^+e^-$ & WWA      &     &          & $5/3$ & 135 & 185 & 170 & 210 \\
\hline
LINAC & $\gamma   \gamma  $
    & 500 GeV & $10 fb^{-1}$        & $1/3$ & 160 & 180 & 175 & 190 \\
\cline{5-9}
$e^+e^-$  & Compton &     &          & $5/3$ & 200 & 205 & 200 & 205 \\
\hline
LINAC & $\gamma^* \gamma^*$
    & 1 TeV & $10 fb^{-1}$        & $1/3$ & 140 & 195 & 285 & 345 \\
\cline{5-9}
$e^+e^-$  & WWA &         &          & $5/3$ & 220 & 325 & 435 & 470 \\
\hline
LINAC & $\gamma \gamma$
    & 1    TeV & $10 fb^{-1}$        & $1/3$ & 300 & 340 & 390 & 405 \\
\cline{5-9}
$e^+e^-$  & Compton &     &          & $5/3$ & 400 & 405 & 410 & 410 \\
\hline \hline
\end{tabular}
\end{center}


\vspace{7mm}
\noindent
{\sf
Table~2:~Accessible mass ranges for leptoquark
pair production~(GeV) for $M_{S,V} \geq 45 \GeV$.
For the case of vector leptoquarks the mass
ranges correspond to $\kappa_G = 1.3, \lambda_G = -0.21$ at the
TEVATRON,
and the minimum vector coupling
$\kappa_G = 1, \lambda_G = 0$ for all other cases.
}

\vspace{3mm}
\noindent
In the foreseeable future, the widest mass range can be explored at LHC
reaching masses   of $O(1~{\rm to}~1.5  \TeV)$
at ${\cal L} = 10~fb^{-1}$.
The accessible mass ranges at future $\gamma \gamma$ colliders operating
at $\sqrt{s} = 1 \TeV$, using laser back scattering to form the photon
beams, reach masses of $O(400 \GeV)$ at the same integrated luminosity.
For comparison, the mass range of $O(200~\GeV)$
and $O(450~\GeV)$, respectively, can be reached for the case of all
scalar and vector leptoquarks discussed in \cite{BRW} for the
pair production process $e^+e^- \rightarrow \Phi \overline{\Phi}$
at $\sqrt{s} = 1 \TeV$ at an integrated luminosity of
${\cal L} = 10~fb^{-1}$, assuming the
minimal  coupling for the vector states,~cf.~\cite{BR}.
The mass ranges of  $O(200 \GeV)$ are accessible at $e^+e^-$ linear
colliders with a cms energy of $\sqrt{s} = 500 \GeV$, a possible
future $ep$ collider at LEP$\otimes$LHC, and at the
TEVATRON. One should
note, however, that the different scattering processes at
$pp$, $p\overline{p}$, $ep$, $e^+e^-$, and $\gamma \gamma$
colliders are not  suited
to limit all the specific production channels equally well.
Particularly the search for
third
generation leptoquarks is difficult
at hadron colliders~\cite{PDG}.
Though the accessible mass ranges for LEP~2~\cite{BR,BBK96A} and HERA
are bounded to $90~\GeV$ and below, depending on the
respective leptoquark type, at both colliders one may search for
these particles.

\section{Conclusions}
\label{sectco}

\vspace{1mm}
\noindent
The differential and integral hadronic pair production cross sections
for scalar and vector leptoquarks were calculated. In the latter case,
we accounted for general anomalous couplings, $\kappa_G$ and $\lambda_G$.
Predictions were made on the discovery
potential of leptoquarks at present and future high energy colliders.
The processes  considered  set mass bounds in a widely model
independent way since the scattering cross sections depend on the known
gauge couplings. In the case of vector leptoquarks in the mass range
$M_V \lsim 1 \TeV$, anomalous couplings also
emerge since these particles
are not gauge bosons. However, as was shown explicitly for the case of
$p\overline{p}$ and $pp$ scattering, a set of anomalous couplings
$\kappa_G^{min}, \lambda_G^{min}$ exists yielding a non-zero, minimal
production cross section. Due to this global mass bounds may be derived.

Hadron colliders cover the widest mass range for leptoquark searches
based on pair production. Therefore the best constraints are expected
from the
TEVATRON in the near future, and later from LHC for those
signatures which can be well separated from the background at these
colliders. It appears to be likely that one may search for first
and second
generation leptoquarks in this way in the mass range up to $1.2$ and
$1.5~\TeV$ for the expected scalar and vector states.

A more difficult  case concerns the 3rd generation leptoquarks decaying
into
$\tau b$, e.g. This type of signature may be more easily    isolated
at colliders with a lower hadronic background. As evident from the
explorable mass ranges, leptoquark searches may be
carried out looking for those spectacular decay channels as
$\Phi \rightarrow t \tau$ at LHC and future linear colliders.

\vspace{4mm}
\noindent
{\bf Acknowledgments:}~We
would like to thank Prof. P. S\"oding for his constant support of the
present project. Our thanks are due to
A.~Pukhov for the
help through the
implementation of the interaction Lagrangian (2) into {\sf CompHEP},
and both A.~Pukhov and V.~Ilyin for discussions.
We would like to thank S. Riemersma for a careful reading of
the manuscript.
E.B. and A.K.
would like to thank DESY--Zeuthen for the warm
hospitality extended to them.
The work has been supported in part by the EC grant `Capital Humain et
Mobilite' CHRX--CT92--0004,
by the grant 95-0-6.4-38 of the Center of Natural
Sciences of the State Committee for Higher Education in Russia,
and by the RFBR grants 96-02-18635a and 96-02-19773a.

\newpage
\appendix
\section{Feynman Rules}
\label{sectAPPA}
We use the convention of ref.~\cite{MUTA}. All momenta are incoming.
The propagators of the
gluon-, ghost-, fermion-, and leptoquark fields are:
\begin{eqnarray}
D_{\mu \nu}^{ab, g}(k) & = & \delta^{ab} d^{\mu \nu}(k) \frac{1}{k^2},
\\
\hat{\Delta}^{ab,\hat{g}}(k) &=& - \delta^{ab} \frac{1}{k^2},
\\
G_{ij}(p) &=&  \delta_{ij} \frac{1}{m_q - \PD},\\
D_{ab}^{S}(k) &=& - \delta_{ab} \frac{1}{k^2 - M_S^2}, \\
D_{ab}^{\mu \nu,V}(k) &=& \delta_{ab} \Delta^{\mu \nu}(k)
\frac{1}{k^2 - M_V^2},
\end{eqnarray}
with
\begin{eqnarray}
d_{\mu \nu}(k) &=&  g_{\mu \nu} - (1 - \xi) \frac{k_{\mu}
k_{\nu}}{k^2},\\
\Delta_{\mu \nu}(k) &=&  g_{\mu \nu} -  \frac{k_{\mu} k_{\nu}}{M_V^2}.
\end{eqnarray}
$\xi$ denotes the gauge parameter of the gluon field in $R_{\xi}$ gauges.

The triple vertices are:
\begin{eqnarray}
V^{ggg,a_1a_2a_3}_{\mu_1 \mu_2 \mu_3} &=& - if^{abc} g_s
\widehat{V}_{\mu_1 \mu_2 \mu_3} (k_1, k_2, k_3),
\\
V_{\mu}^{\hat{g}\hat{g}g, a_1a_2a_3} &=& - ig_sf^{abc} k_{2 \mu},
\\
V^{qqg,j i a}_{\mu} &=& g_s \gamma_{\mu} t^a_{ij},
\\
V^{\overline{S}Sg,aij}_{\mu_3} (k_1, k_2, k_3)
&=& g_s(t^a)^{ij}(k_2-k_1)_{\mu_3},
\\
V_{\mu_{1} \mu_{2} \mu_{3}}^{\overline{V} Vg, aij}(k_1, k_2, k_3) &=&
g_s(t^a)^{ij} \bigg[ \widehat{V}_{\mu_{1} \mu_{2} \mu_{3}} +
\kappa_G \widehat{V}_{\mu_{1} \mu_{2} \mu_{3}}^{\kappa}
+ \frac{\lambda_G}{M_V^2} \widehat{V}_{\mu_1 \mu_2 \mu_3}
^{\lambda} \bigg],
\end{eqnarray}
where
\begin{eqnarray}
\widehat{V}_{\mu_1, \mu_2, \mu_3}(k_1, k_2, k_3) &=&
  (k_1 - k_2)_{\mu_3} g_{\mu_1 \mu_2}
+ (k_2 - k_3)_{\mu_1} g_{\mu_2 \mu_3}
+ (k_3 - k_1)_{\mu_2} g_{\mu_3 \mu_1},   \\
\widehat{V}_{\mu_{1} \mu_{2} \mu_{3}}^{\kappa} (k_1, k_2, k_3) &=&
k_{3 \mu_1} g_{\mu_2 \mu_3} - k_{3 \mu_2} g_{\mu_1 \mu_3},
\\
\widehat{V}^{\lambda}_{\mu_1 \mu_2 \mu_3}(k_1, k_2, k_3) & = &
(k_1. k_2) (k_{3 \mu_1} g_{\mu_2 \mu_3} -  k_{3 \mu_2} g_{\mu_1
  \mu_3})
+ (k_2. k_3) (k_{1 \mu_2} g_{\mu_1 \mu_3} -  k_{1 \mu_3} g_{\mu_1
\mu_2}) \nonumber\\
& + & (k_3. k_1) (k_{2 \mu_3} g_{\mu_1 \mu_2} -  k_{2 \mu_1}
g_{\mu_2 \mu_3})
+ k_{1 \mu_3} k_{2 \mu_1} k_{3 \mu_2} - k_{1 \mu_2}
k_{2 \mu_3} k_{3 \mu_1}.
\end{eqnarray}
The four--vertices are:
\begin{eqnarray}
W_{\mu_1\mu_2\mu_3\mu_4}^{gggg,a_1a_2a_3a_4}(k_1,k_2,k_3,k_4)
&=& -g_s^2 \Bigg \{
f^{a_1 a_2 b} f^{a_3 a_4 b} ( g_{\mu_1 \mu_3} g_{\mu_2 \mu_4}
                       - g_{\mu_1 \mu_4} g_{\mu_2 \mu_3}) \nonumber\\
&+& f^{a_1 a_3 b} f^{a_2 a_4 b} ( g_{\mu_1 \mu_2} g_{\mu_3 \mu_4}
                       - g_{\mu_1 \mu_4} g_{\mu_2 \mu_3}) \nonumber\\
&+& f^{a_1 a_4 b} f^{a_3 a_2 b} ( g_{\mu_1 \mu_3} g_{\mu_2 \mu_4}
                       - g_{\mu_1 \mu_2} g_{\mu_3 \mu_4}) \Bigg \},
\\
W^{\overline{S}Sgg, ija_1a_2}_{\mu_1 \mu_2}(p_1, p_2, p_3, p_4)
&=& g^2_s (
t^{a_1}t^{a_2} + t^{a_2}t^{a_1})^{ij} g_{\mu_1 \mu_2},
\\
W^{\overline{V}Vgg, ija_1a_2}_{\mu_1 \mu_2 \mu_3 \mu_4}
(p_1, p_2, p_3, p_4)  &=& - g^2_s \bigg\{ (t^{a_1}t^{a_2})^{ij}
\bigg[\widehat{W}_{\mu_1 \mu_2 \mu_3 \mu_4}
+ \kappa_G \widehat{W}^{\kappa}_{\mu_1 \mu_2 \mu_3 \mu_4}
+ \frac{\lambda_G}{M_V^2}
\widehat{W}^{\lambda}_{\mu_1 \mu_2 \mu_3 \mu_4} \bigg]
\nonumber\\
&+& ( t^{a_2}t^{a_1})^{ij} \Big[ \widehat{W}_{\mu_1 \mu_2
  \mu_4 \mu_3} +  \kappa_G
\widehat{W}_{\mu_1 \mu_2 \mu_4 \mu_3}^{\kappa}
+ \frac{\lambda_G}{M_V^2} \widehat{W}^{\lambda}_{\mu_1 \mu_2
  \mu_4 \mu_3} \Big] \bigg\},
\end{eqnarray}
with
\begin{eqnarray}
\widehat{W}_{\mu_1 \mu_2 \mu_3 \mu_4}(p_1, p_2, p_3, p_4)
&=& g_{\mu_1 \mu_2}
g_{\mu_3 \mu_4} + g_{\mu_1 \mu_3} g_{\mu_2 \mu_4} - 2g_{\mu_1 \mu_4}
g_{\mu_2 \mu_3},
\\
\widehat{W}_{\mu_1 \mu_2 \mu_3 \mu_4}^{\kappa}(p_1, p_2, p_3, p_4)
&=&g_{\mu_1\mu_4} g_{\mu_2 \mu_3}
- g_{\mu_1 \mu_3} g_{\mu_2 \mu_4},
\\
\widehat{W}_{\mu_1 \mu_2 \mu_3 \mu_4}^{\lambda}(p_1, p_2,
p_3, p_4) &=& (p_1.p_2)(g_{\mu_1 \mu_4} g_{\mu_2 \mu_3} -
g_{\mu_1 \mu_3} g_{\mu_2  \mu_4}),
\nonumber\\
&+& (p_1.p_3)(g_{\mu_1 \mu_4} g_{\mu_2 \mu_3}
            - g_{\mu_1 \mu_2} g_{\mu_3 \mu_4})
\nonumber\\
&+& (p_2.p_4)(g_{\mu_1 \mu_4} g_{\mu_2 \mu_3}
- g_{\mu_1 \mu_2} g_{\mu_3 \mu_4})
\nonumber\\
&+& g_{\mu_1 \mu_2}(p_{1 \mu_3} p_{3 \mu_4} + p_{2 \mu_4} p_{4 \mu_3} +
p_{1 \mu_4} p_{2 \mu_3} - p_{1 \mu_3} p_{2 \mu_4})
\nonumber\\
&+& g_{\mu_1 \mu_3}(p_{1 \mu_2} p_{2 \mu_4} + p_{1 \mu_4}
p_{3 \mu_2} - p_{1 \mu_2} p_{3 \mu_4})
\nonumber\\
&-& g_{\mu_1 \mu_4}(p_{1 \mu_2}
p_{2 \mu_3} + p_{1 \mu_3} p_{3 \mu_2} + p_{2 \mu_3} p_{4 \mu_2})
\nonumber\\
&-& g_{\mu_2 \mu_3} (p_{2 \mu_1} p_{1 \mu_4} +
p_{2 \mu_4} p_{4 \mu_1} + p_{1 \mu_4} p_{3 \mu_1})
\nonumber\\
&+& g_{\mu_2 \mu_4}(p_{2 \mu_1} p_{1 \mu_3}
+ p_{2 \mu_3}p_{4 \mu_1} - p_{2 \mu_1} p_{4 \mu_3})
\nonumber\\
&+& g_{\mu_3 \mu_4} ( p_{1 \mu_2}p_{3 \mu_1}
+ p_{2 \mu_1} p_{4 \mu_2}).
\end{eqnarray}
\newpage
\section{Coefficients of the production cross section of vector
leptoquarks}
\label{sectAPPB}
The functions $F_i(\hat{s},\beta,\cos \theta)$  which
determine the differential pair production cross section for
$gg \rightarrow V\overline{V}$
are:

\begin{eqnarray}
F_0 &=& \bigg[ 19 - 6 \beta^2 +  6 \beta^4 + \left( 16 - 6 \beta^2
\right) \beta^2 \cos^2 \theta + 3 \CBB \bigg] \cdot
\left( 7+9 \beta^2 \cos^2 \theta \right)
\\
F_1 &=& -4 \cdot \left( 77+143 \beta^2 \cos^2 \theta + 36 \beta^4
\cos^4 \theta \right)
\\
F_2 &=& -8 \cdot \left( 7+11 \CB - 18 \CBB \right)
\\
F_3  &=& 2 \cdot \left( 117+185 \CB + 18 \CBB \right)
 + 2 \SM \left( 8- \CB - 7\CBB \right)
\nonumber\\
&+& \frac{7}{4} \SMM \BR ^2
\\
F_4 &=& -4 \cdot \left( 19+27 \CB + 18 \CBB \right) +10 \SM \BR
\left( 7-\CB \right)
\\
F_5 &=& 2 \cdot \left( 19+27 \CB + 18 \CBB \right) - \SM \BR
\left( 65+29 \CB \right)
\nonumber\\
&+& \frac{1}{8} \SMM \BR \left( 97+2 \CB -115 \CBB \right)
+ \SMMM \frac{9}{4} \BR^3
\\
F_6 &=& -61-67 \CB - \frac{1}{2} \SM \BR \left( 39+14 \CB \right)
\nonumber\\
&-& \frac{7}{4} \SMM \BR ^2
\\
F_7 &=& 127+129 \CB -  \frac{1}{2} \SM \BR \left( 89+3 \CB \right)
\nonumber\\
&+& \frac{1}{4} \SMM \BR ^2 \left( -23+18 \CB \right)
\\
F_8 &=& -71-57 \CB + \frac{1}{2} \SM \BR \left( 170+21 \CB \right)
\nonumber\\
&+& \frac{1}{4} \SMM \BR \left( -59+40 \CB + 27 \CBB \right)
- \frac{9}{4} \SMMM \BR^3
\\
F_9 &=& 5 \BR - \SM \BR \left( 21+2 \CB \right)
\nonumber\\
&+& \frac{1}{4} \SMM \BR^2 \left( 74+9 \CB \right)
\nonumber\\
&+& \frac{1}{4}\SMMM \BR^2 \left( -15+8 \CB \right)
\\
F_{10} &=& 3+5 \CB +  \frac{5}{4} \SM \BR \left( 4- \CB \right)
\nonumber\\
&+& \frac{1}{32} \SMM \BR^2 \left( 25+13 \CB \right)
\\
F_{11} &=& -4 \cdot \left( 3+5 \CB \right) -5 \SM \BR^2
\nonumber\\
&+& \frac{1}{8} \SMM \BR^2 \left( 35-13 \CB \right)
\\
F_{12} &=& 6 \cdot \left( 3+5 \CB \right) - \frac{15}{2} \SM \BR
\left( 2+ \CB \right)
\nonumber\\
&+& \frac{1}{16} \SMM \BR \left( -23+54 \CB -39 \CBB \right)
\nonumber\\
&+& \frac{1}{64} \SMMM \BR^2 \left( 113-49 \CB \right)
\\
F_{13} &=& -4 \cdot \left( 3+5 \CB \right) + 5 \SM \BR
\left( 5+ \CB \right)
\nonumber\\
&-&  \frac{1}{8} \SMM \BR^2 \left( 119+13 \CB \right)
\nonumber\\
&+& \frac{1}{32} \SMMM \BR^2 \left( 79-15 \CB \right)
\\
F_{14} &=& 3+5 \CB - \frac{5}{4} \SM \BR
\left( 8+ \CB \right)
\nonumber\\
&+& \frac{1}{32} \SMM \BR \left( 321  - 324 \CB - 13 \CBB \right)
\nonumber\\
&+& \frac{11}{64} \SMMM \BR^2 \left( -23+7 \CB \right)
\nonumber\\
&+& \frac{1}{256} \SMMMM \BR^2 \left( 135-22\CB+15 \CBB \right).
\end{eqnarray}
The coefficients $\widetilde{F}_i(\hat{s},\beta)$ for the integrated
cross section for $gg \rightarrow V\overline{V}$ are:
\\
\begin{eqnarray}
\widetilde{F}_0 &=& \B \left( \frac{523}{4} - 90 \beta^2
+ \frac{93}{4} \beta^4 \right)
- \frac{3}{4} (65 - 83 \beta^2 + 19 \beta^4 - \beta^6) \XL
\nonumber\\
\widetilde{F}_1 &=&
 -4 \beta(41 - 9\beta^2) - \frac{87}{2} (1 - \beta^2) \XL
\\
\widetilde{F}_2 &=&
 36 \beta (1 - \beta^2) - 25(1 - \beta^2) \XL
\\
\widetilde{F}_3 &=& \beta(75 - 9 \beta^2) + \frac{7}{4} \beta
\SM - \frac{1}{4}
(1 - 61 \beta^2) \XL
\\
\widetilde{F}_4 &=& -2 \beta(20 - 9 \beta^2) + \frac{1}{2}
(91 - 31 \beta^2) \XL
\\
\widetilde{F}_5 &=& \beta \left( \frac{209}{6} - 9 \beta^2 \right)
+ \frac{263}{12} \beta \SM + \frac{3}{2} \beta
\SMM - \left( \frac{219}{4} - \frac{31}{4}
\beta^2 + \SM \right) \XL
\\
\widetilde{F}_6 &=& -9 \beta - \frac{7}{4} \beta \SM
- \left( \frac{103}{8} + \frac{3}{8} \beta^2 \right) \XL
\\
\widetilde{F}_7 &=& \frac{55}{2} \beta - \frac{17}{4} \beta
\SM - \left( \frac{185}{8} - \frac{1}{8} \beta^2
\right) \XL
\\
\widetilde{F}_8 &=& -\frac{35}{2} \beta - 22 \beta \SM
- \frac{3}{2} \beta \SMM +
\left( \frac{375}{8} + \frac{7}{8} \beta^2 + \SM
\right) \XL
\\
\widetilde{F}_9 &=& - \beta + \frac{199}{12} \beta \SM
- \frac{37}{12} \beta \SMM - \left( \frac{87}{8}
+ \frac{5}{8} \beta^2 \right) \XL
\\
\widetilde{F}_{10} &=& \frac{41}{24} \beta + \frac{11}{12} \beta
\SM + \left( \frac{7}{4} + \frac{1}{8} \beta^2
\right) \XL
\\
\widetilde{F}_{11} &=& - \frac{41}{6} \beta
 + \frac{23}{6} \beta
\SM + \frac{1}{2} (1 - \beta^2) \XL
\\
\widetilde{F}_{12} &=& \frac{41}{4} \beta + \frac{43}{48} \beta
\SM + \frac{145}{96} \beta \SMM -
\left( 12- \frac{3}{4} \beta^2 + \frac{1}{4}\frac{\hat{s}}{M^2_V}
\right) \XL
\\
\widetilde{F}_{13} &=& - \frac{41}{6} \beta - \frac{355}{24} \beta
\frac{\hat{s}}{M^2_V} + \frac{37}{16} \beta \SMM
+ \left( \frac{31}{2} - \frac{1}{2} \beta^2 \right) \XL
\\
\widetilde{F}_{14} &=& \frac{41}{24} \beta + \frac{37}{4} \beta
\SM - \frac{113}{32} \beta \SMM
+ \frac{49}{96} \beta \SMMM - \left(
\frac{23}{4} - \frac{1}{8} \beta^2 + \frac{1}{4} \SM
\right) \XL .
\\
\end{eqnarray}
Finally, the coefficients for the differential
and the integrated cross section for
$q \overline{q} \rightarrow V\overline{V}$,
$G_i(\hat{s},\beta,\cos \theta)$ and
$\widetilde{G}_i(\hat{s},\beta)$, are given by
\\
\begin{eqnarray}
G_0 &=& 1 + \frac{1}{16} \left [\SM - (1 + 3 \B^2) \right ] \st \\
G_1 &=& -1 - \frac{1}{8} \left [\SM - 2 \right ] \st \\
G_2 &=& 1   \\
G_3 &=& \frac{1}{4} + \frac{1}{16} \left [ \SM - 2 \right ] \st \\
G_4 &=& - \frac{1}{2} + \frac{1}{4} \st \\
G_5 &=& \frac{1}{4} + \frac{1}{8} \left [ \SM - 1 \right ] \st  \\
\widetilde{G}_0 &=& \frac{1}{24} \SM + \frac{23 - 3 \B^2}{24} \\
\widetilde{G}_1 &=& - \frac{1}{12} \SM - \frac{5}{6} \\
\widetilde{G}_2 &=& 1 \\
\widetilde{G}_3 &=& \frac{1}{24} \SM + \frac{1}{6} \\
\widetilde{G}_4 &=& - \frac{1}{3} \\
\widetilde{G}_5 &=& \frac{1}{12} \SM + \frac{1}{6}.
\end{eqnarray}

%

\newpage

\vspace{8cm}
\begin{picture}(120,20)(100,100) \centering
\put(135,-300){\epsfig{file=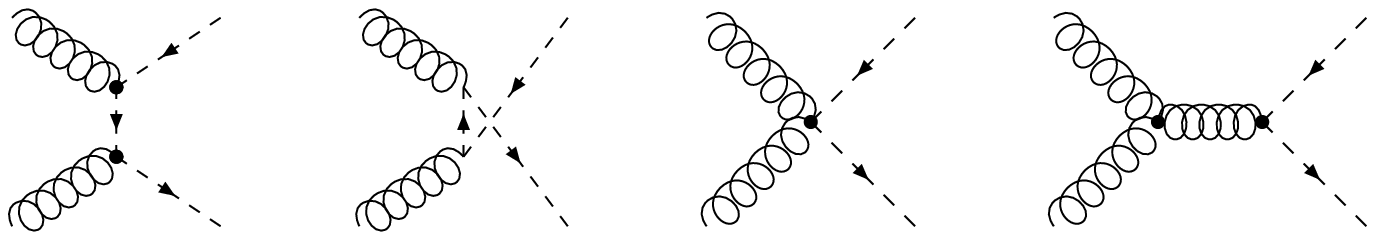,width=16cm}}
\end{picture}

\vspace{8cm}
\noindent
\small
{\sf
Figure~1:
Diagrams describing leptoquark pair production via gluon--gluon fusion.
Here
the dashed lines denote  both scalar and vector leptoquarks.}

\normalsize

\vspace{5cm}
\begin{picture}(120,20)(100,100) \centering
\put(135,-150){\epsfig{file=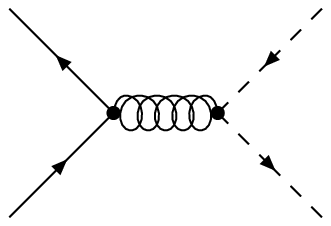,width=16cm}}
\end{picture}

\vspace{2cm}
\noindent
\small
\begin{center}
{\sf
Figure~2:
Diagram for the subprocess $q \overline{q} \rightarrow \Phi_{S,V}
\overline{\Phi}_{S,V}$.}
\end{center}
\normalsize
\newpage
\begin{center}

\mbox{\epsfig{file=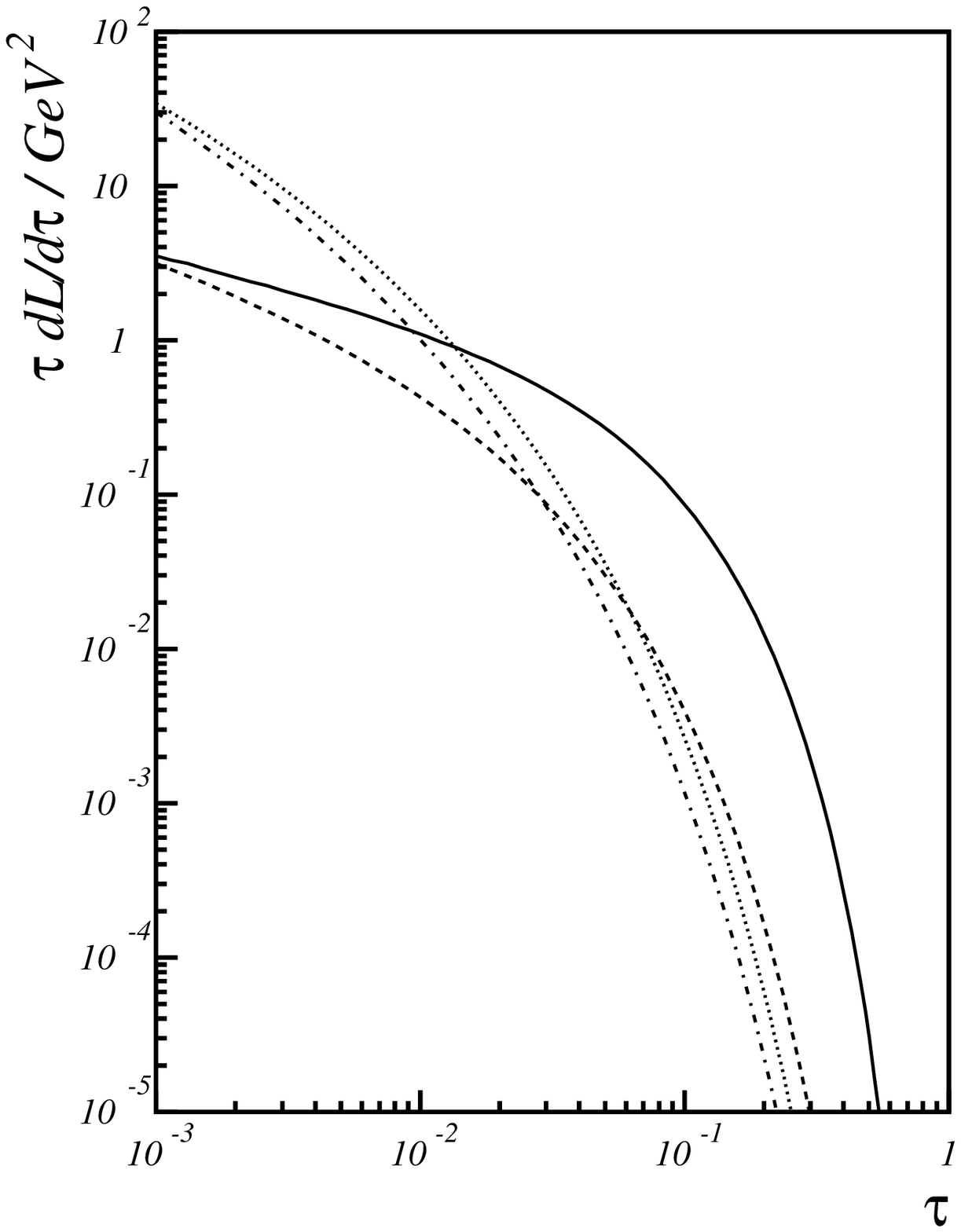,height=18cm,width=16cm}}

\vspace{2mm}
\noindent
\small
\end{center}
{\sf
Figure~3~:~Parton luminosities $\tau d {\cal L}^{q,g}/d \tau$
for different hadron colliders using
the
CTEQ3 parametrization~\cite{CTEQ3}.
Full line: quark-antiquark luminosity
for $p\overline{p}$ at $\sqrt{S} = 1.8 \TeV$;
dashed line: quark-antiquark luminosity for $pp$ at $\sqrt{S} = 14 \TeV$;
dotted line: gluon-gluon luminosity
for $p\overline{p}$ at $\sqrt{S} = 1.8 \TeV$;
dash-dotted
line: gluon-gluon luminosity for $pp$ at $\sqrt{S} = 14 \TeV$.
}
\normalsize
\newpage
\begin{center}

\mbox{\epsfig{file=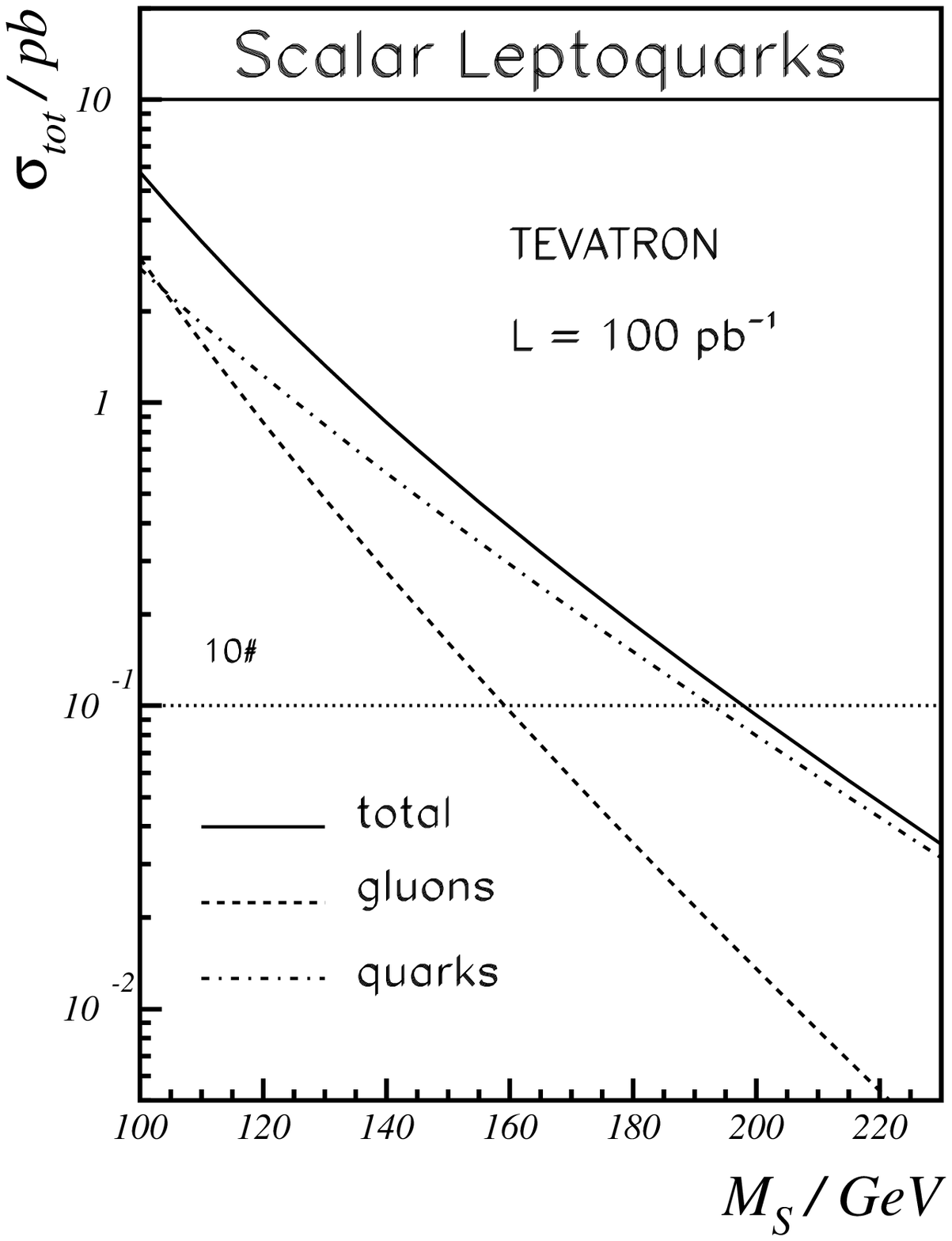,height=18cm,width=16cm}}

\vspace{2mm}
\noindent
\small
\end{center}
{\sf
Figure~4a:~Integrated cross
sections for scalar leptoquark pair production at the
TEVATRON,
$\sqrt{S}~=~1.8~\TeV$.}
\normalsize
\newpage
\begin{center}

\mbox{\epsfig{file=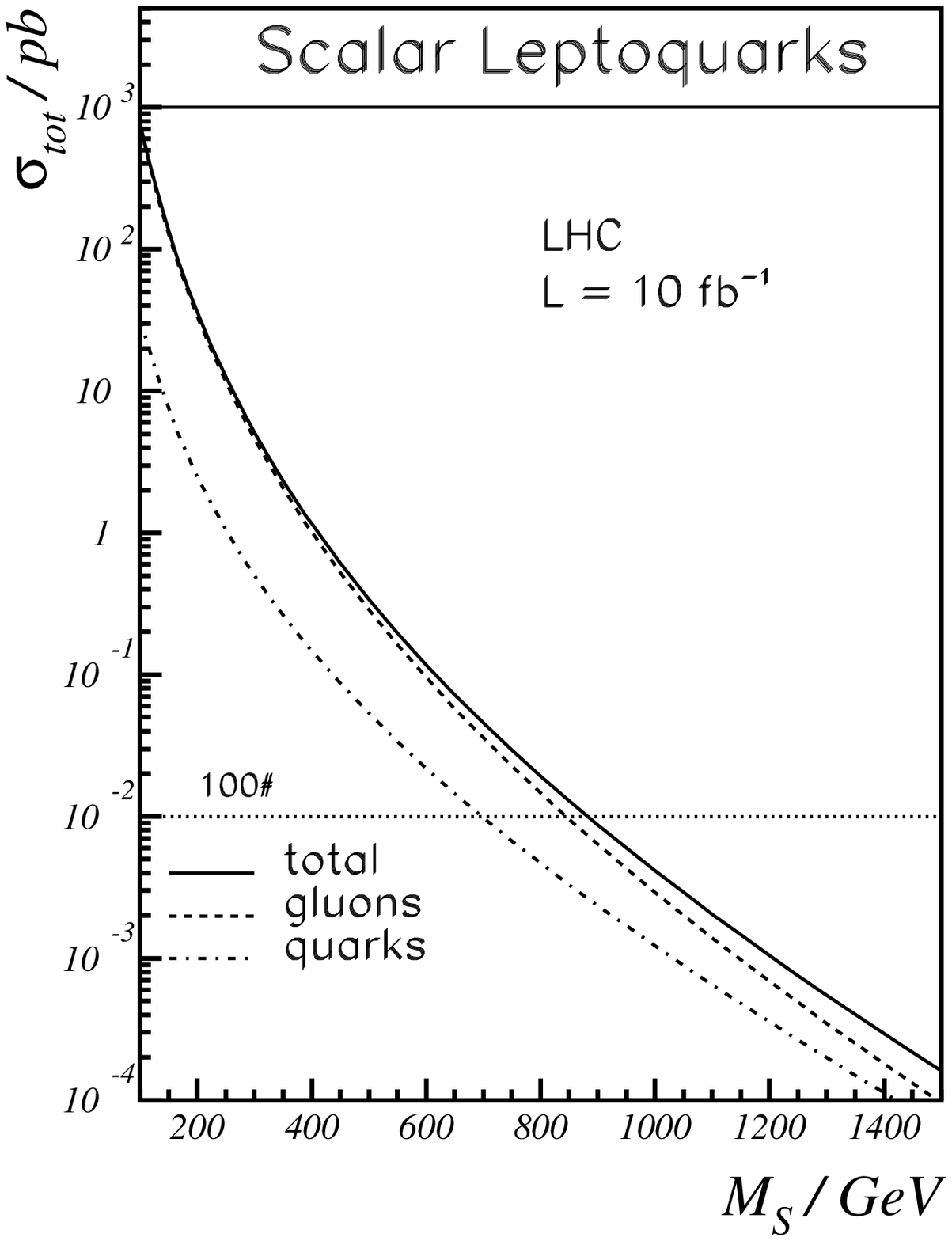,height=18cm,width=16cm}}

\vspace{2mm}
\noindent
\small
\normalsize
\end{center}
{\sf
Figure~4b:~Integrated cross
sections for scalar leptoquark pair production at LHC,
$\sqrt{S}~=~14~\TeV$.}
\newpage
\begin{center}

\mbox{\epsfig{file=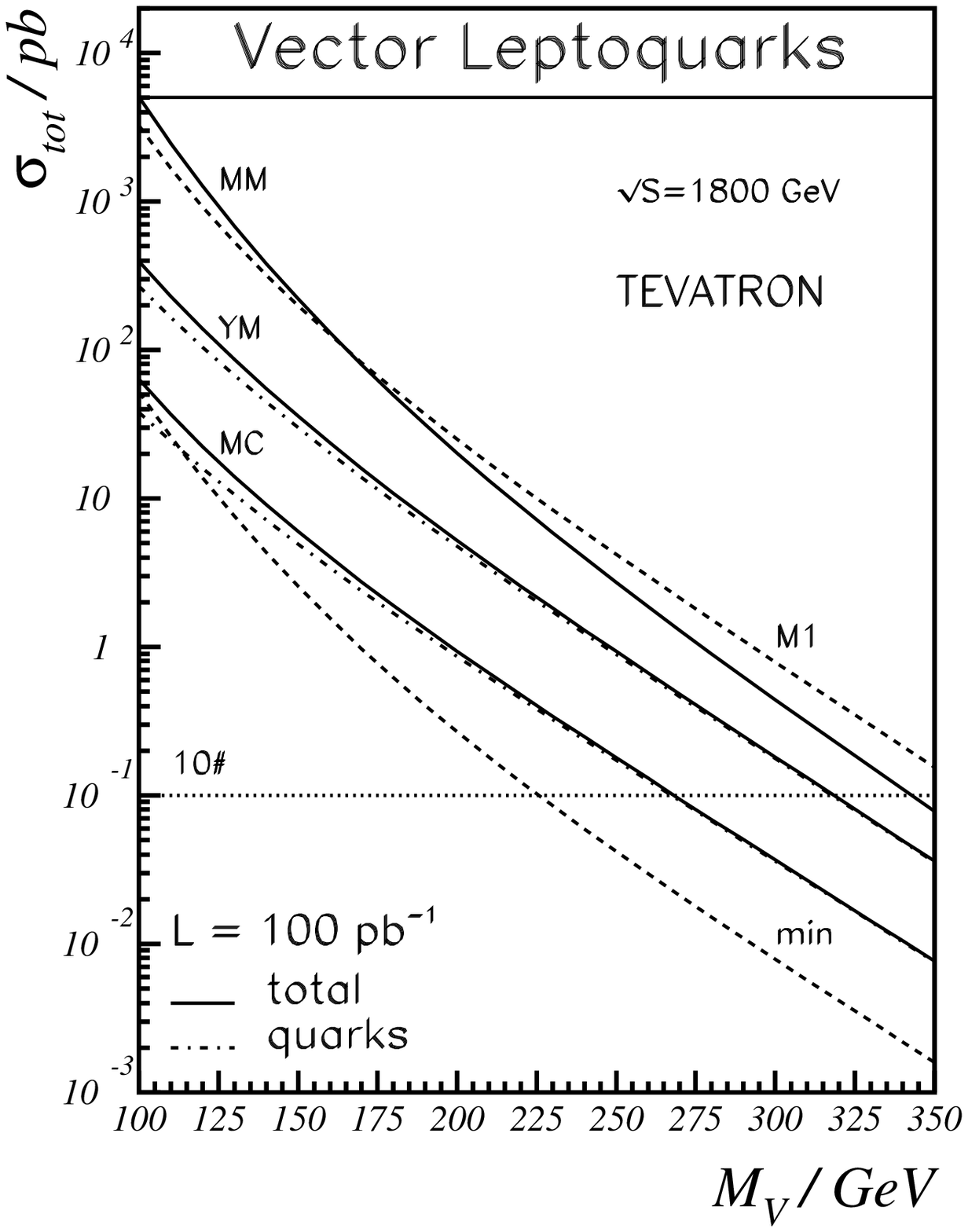,height=18cm,width=16cm}}

\vspace{2mm}
\noindent
\small
\end{center}
{\sf
Figure~5a:~Integrated cross
sections for vector leptoquark pair production at the
TEVATRON,
$\sqrt{S}~=~1.8~\TeV$.
The quark  contributions for
the cases of a Yang-Mills type coupling (YM),
$\kappa_G = \lambda_G  \equiv
0$, and the minmal coupling $\kappa_G = 1, \lambda_G = 0$ (MC) are
shown explicitly. The upper full line denotes the total
cross section for the case $\kappa_G = \lambda_G = -1$~(MM),
while the upper dashed line corresponds
to $\kappa_G = -1,\lambda_G = +1$~(M1), and  the lower dashed line
to $\kappa_G = 1.3, \lambda_G = -0.21$~(min).
}
\normalsize
\newpage
\begin{center}

\mbox{\epsfig{file=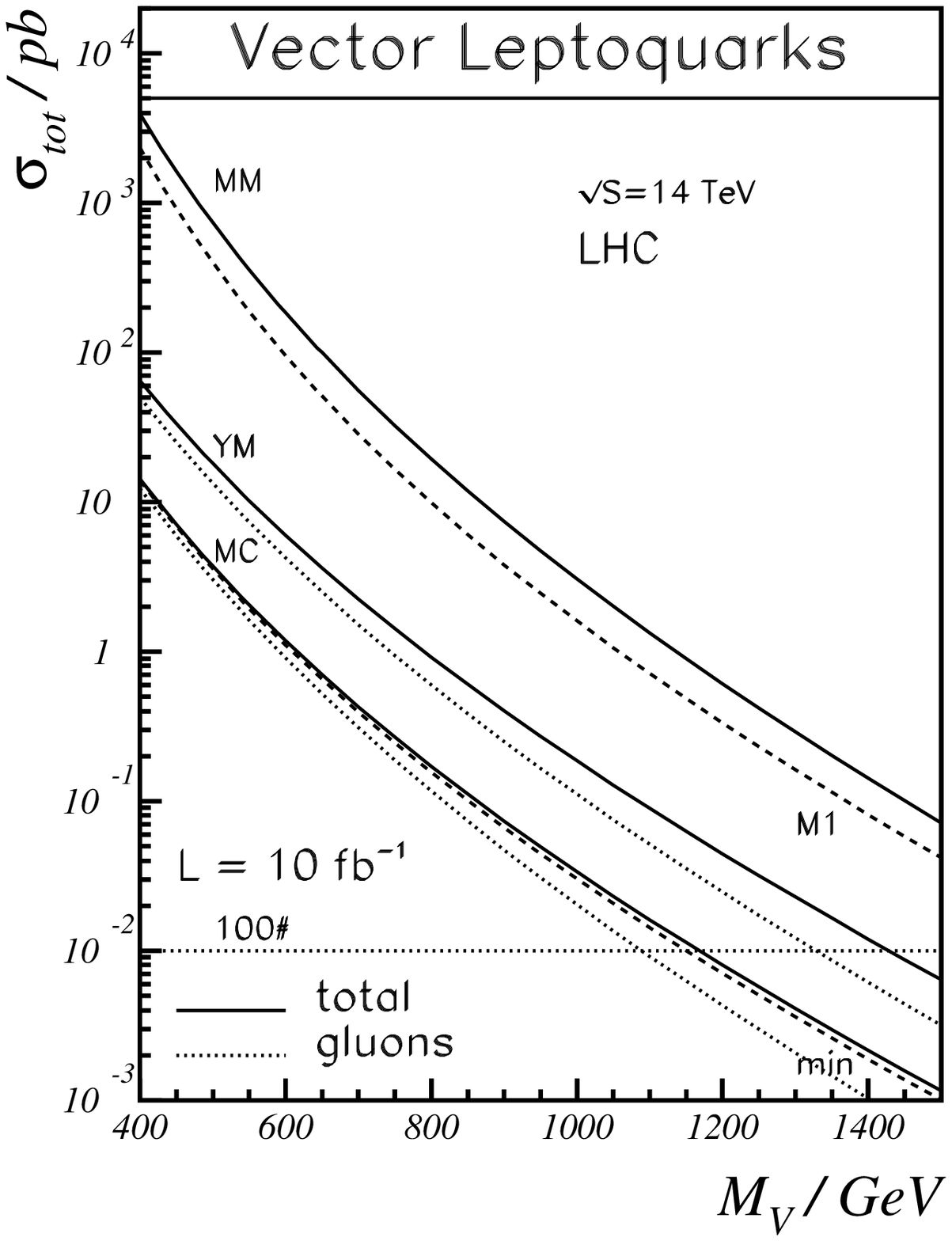,height=18cm,width=16cm}}

\vspace{2mm}
\noindent
\small
\end{center}
{\sf
Figure~5b:~Integrated cross sections for vector leptoquark pair
production at LHC, $\sqrt{S}~=~14~\TeV$. For the Yang-Mills type (YM)
and minimal vector couplings (MC) also the gluon contributions are
shown explicitly (dotted lines).
The lower dashed line  corresponds to
$\kappa_G = 1, \lambda_G = -0.05$~(min).
The other notations are the same as in figure~5a.
}
\normalsize
\newpage
\begin{center}

\mbox{\epsfig{file=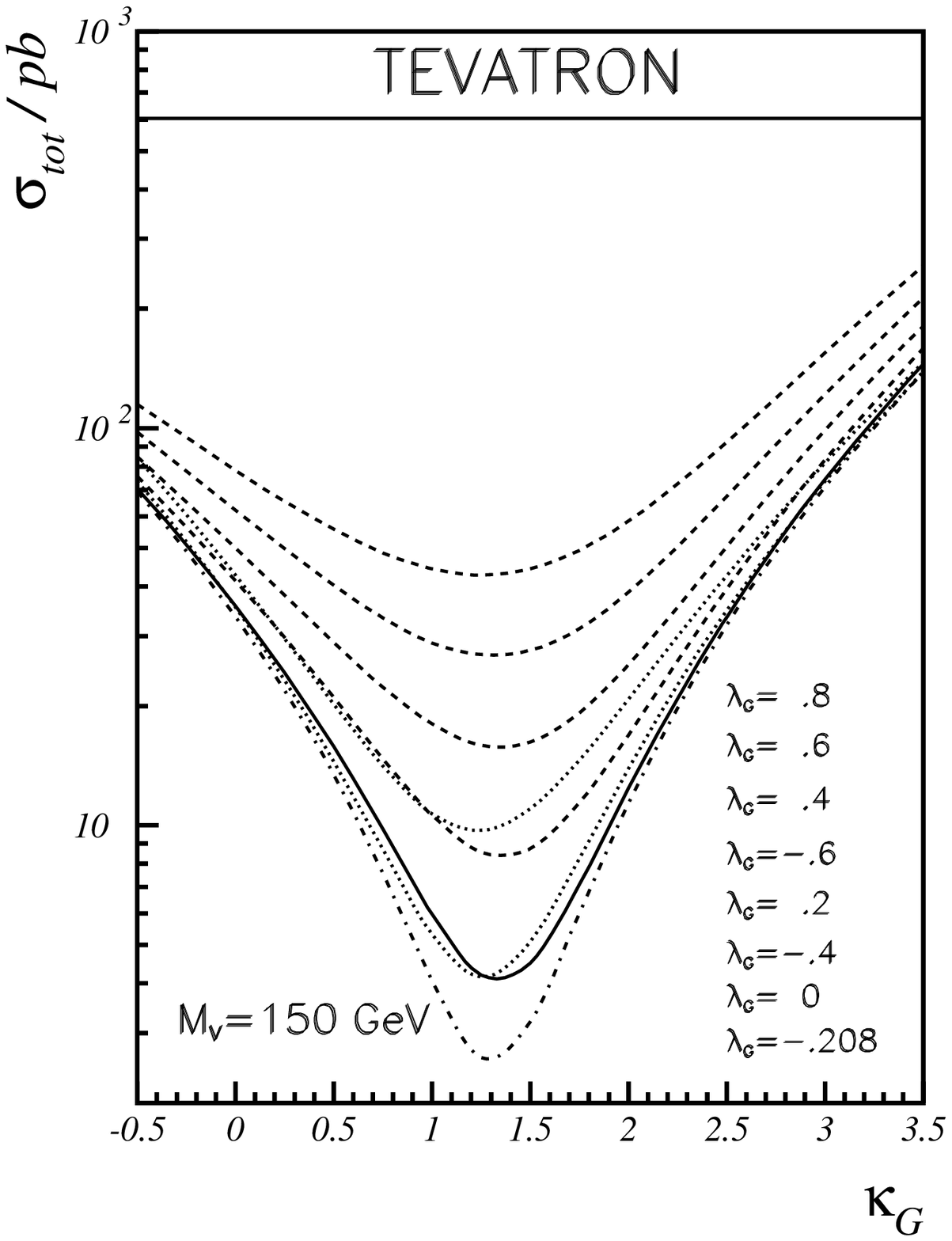,height=18cm,width=16cm}}

\vspace{2mm}
\noindent
\small
\end{center}
{\sf
Figure~6a:~Dependence of the
integrated cross
sections for vector leptoquark pair production on the anomalous
couplings~$\kappa_G$ and $\lambda_G$
at the
TEVATRON,
$\sqrt{S}~=~1.8~\TeV$ for $M_V = 150 \GeV$.
The order of the values of $\lambda_G$ follows
the position of the respective minimum. Dashed lines: $\lambda_G > 0$,
dotted lines: $\lambda_G < 0$, full line: $\lambda_G = 0$,
dash-dotted line:  $\lambda_G = -0.208$.
}
\normalsize
\newpage
\begin{center}

\mbox{\epsfig{file=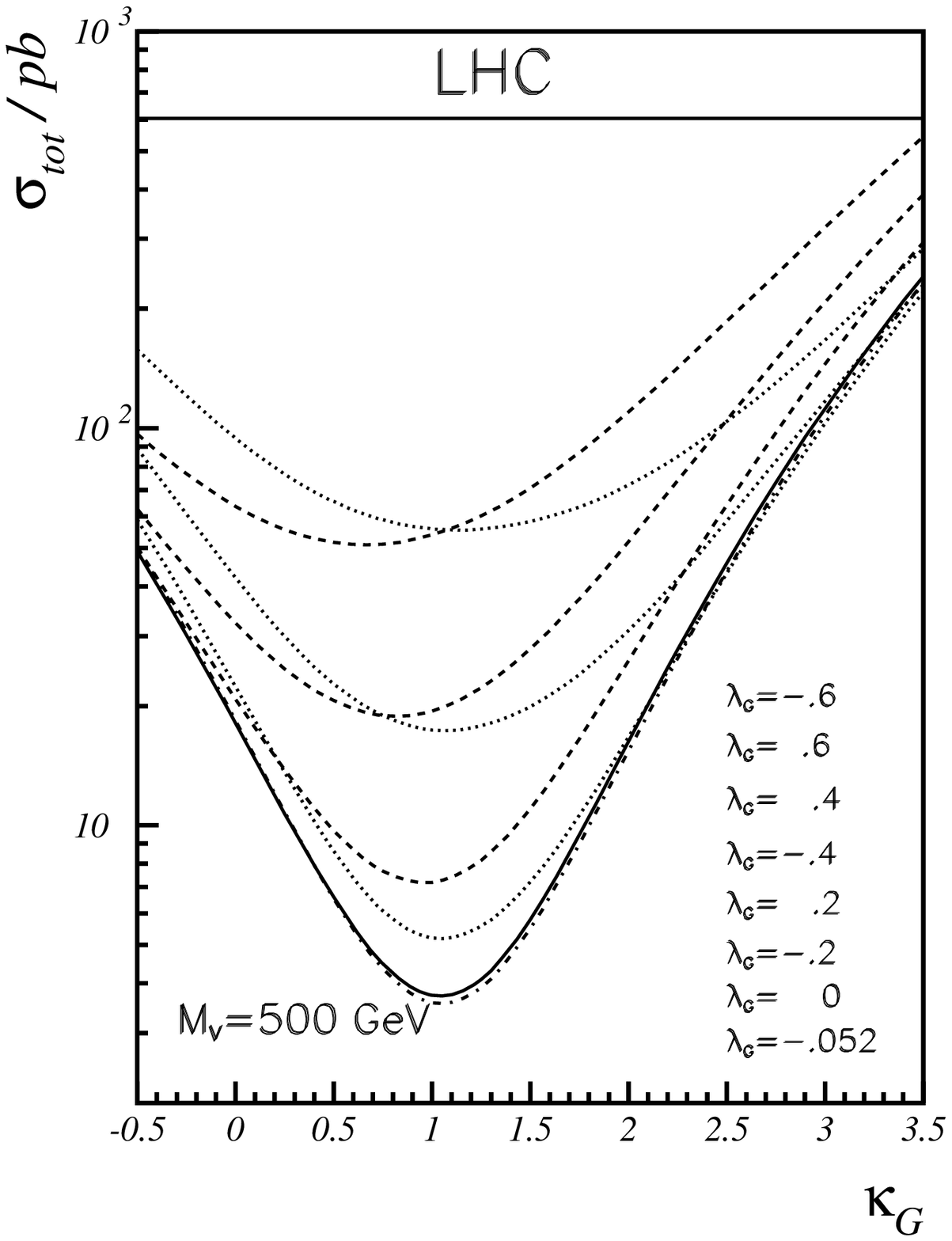,height=18cm,width=16cm}}

\vspace{2mm}
\noindent
\small
\end{center}
{\sf
Figure~6b:~Dependence of the
integrated cross
sections for vector leptoquark pair production on the anomalous
couplings~$\kappa_G$ and $\lambda_G$
at LHC,
$\sqrt{S}~=~14~\TeV$ for $M_V = 500 \GeV$.
The dash-dotted line corresponds to $\lambda_G = -0.052$.
The other parameters are the same as in figure~6a.
}
\normalsize
\newpage
\begin{center}

\mbox{\epsfig{file=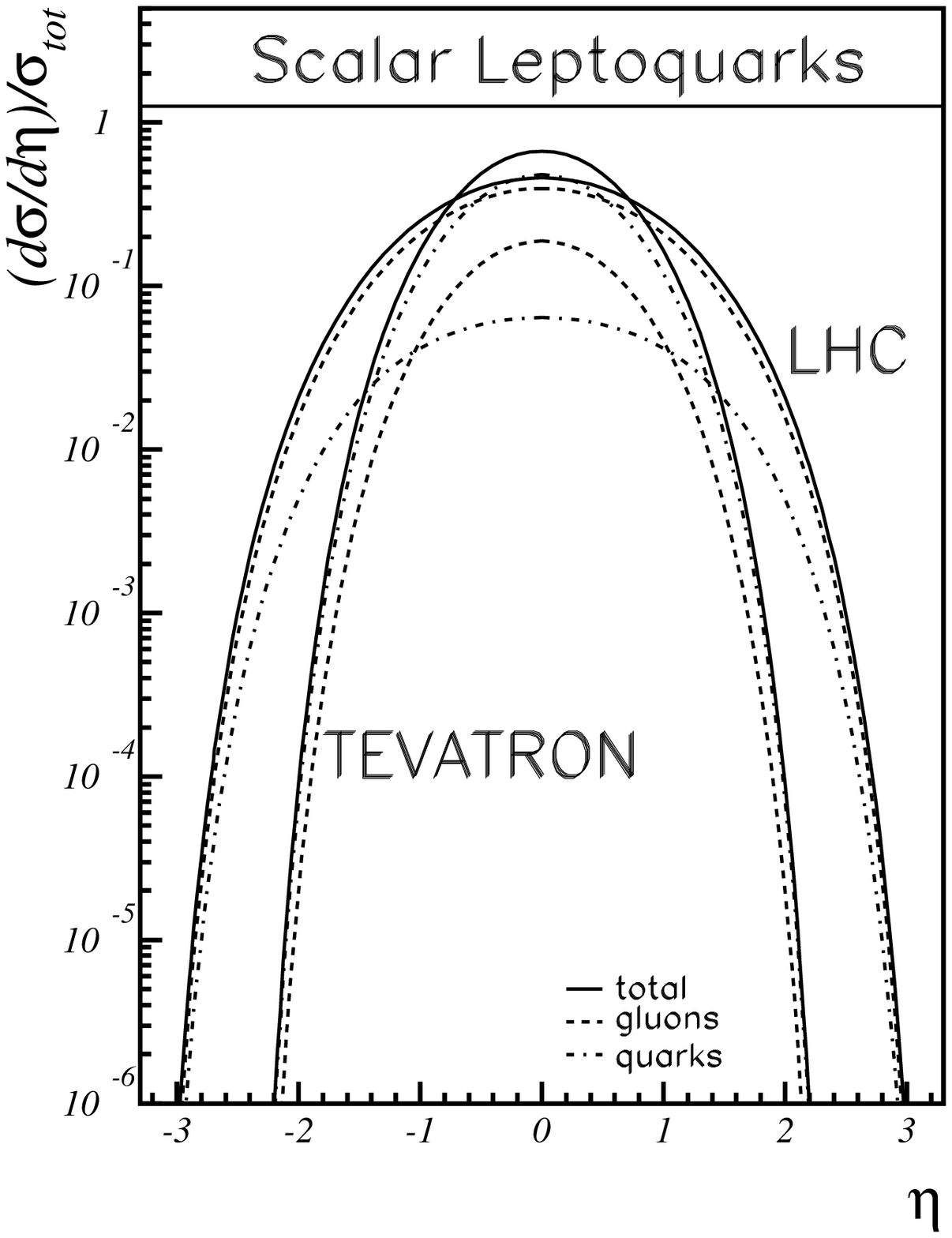,height=18cm,width=16cm}}

\vspace{2mm}
\noindent
\small
\end{center}
{\sf
Figure~7a:~Rapidity distributions for scalar leptoquark pair production
at the
TEVATRON and LHC.
The leptoquark masses are the same as in
figures~6a,b.}
\normalsize
\newpage
\begin{center}

\mbox{\epsfig{file=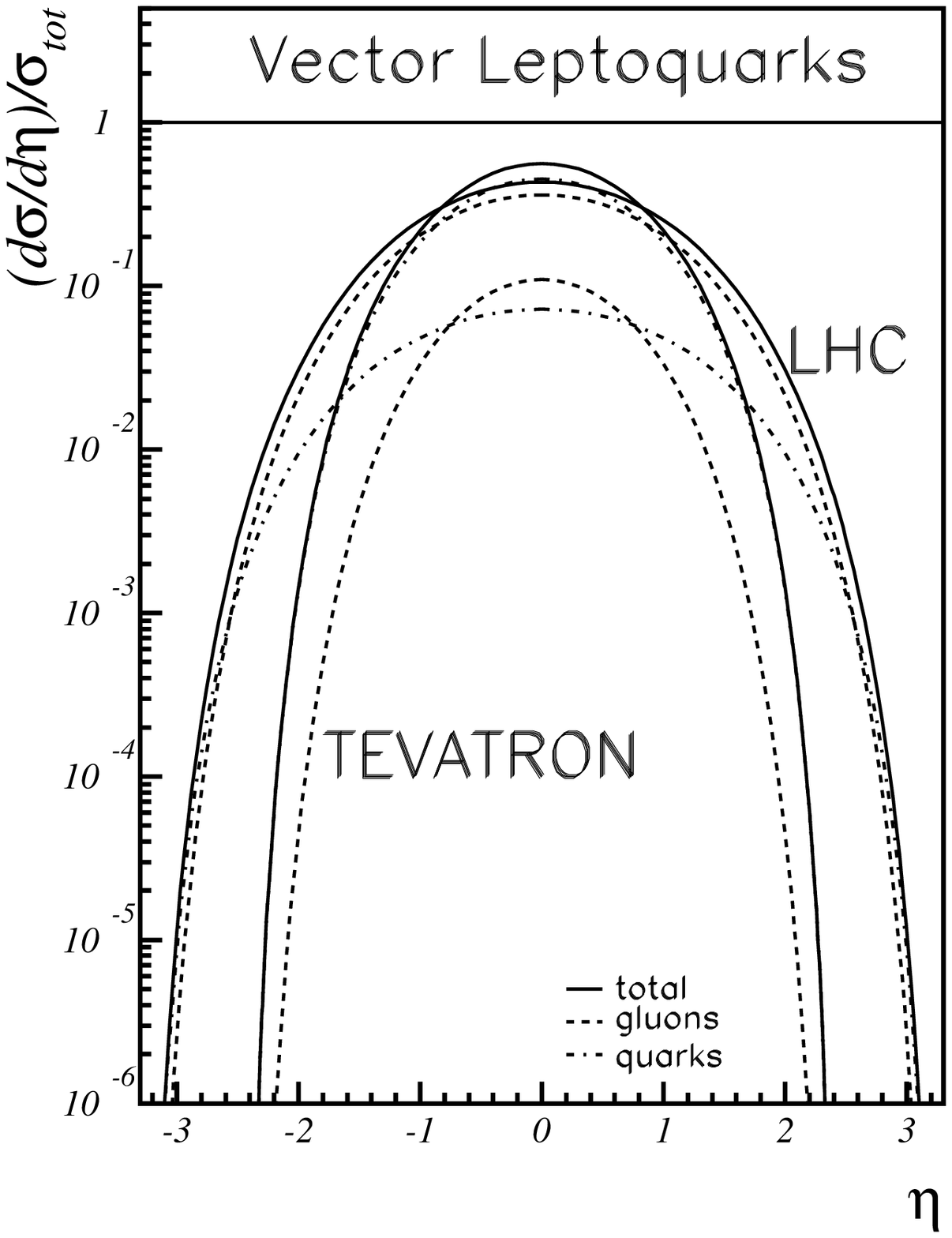,height=18cm,width=16cm}}

\vspace{2mm}
\noindent
\small
\end{center}
{\sf
Figure~7b:~Rapidity distributions for vector leptoquark pair production
at the
TEVATRON and LHC assuming  minimal vector couplings.
The notation is the same as in figure~7a.
}
\normalsize
\newpage
\begin{center}

\mbox{\epsfig{file=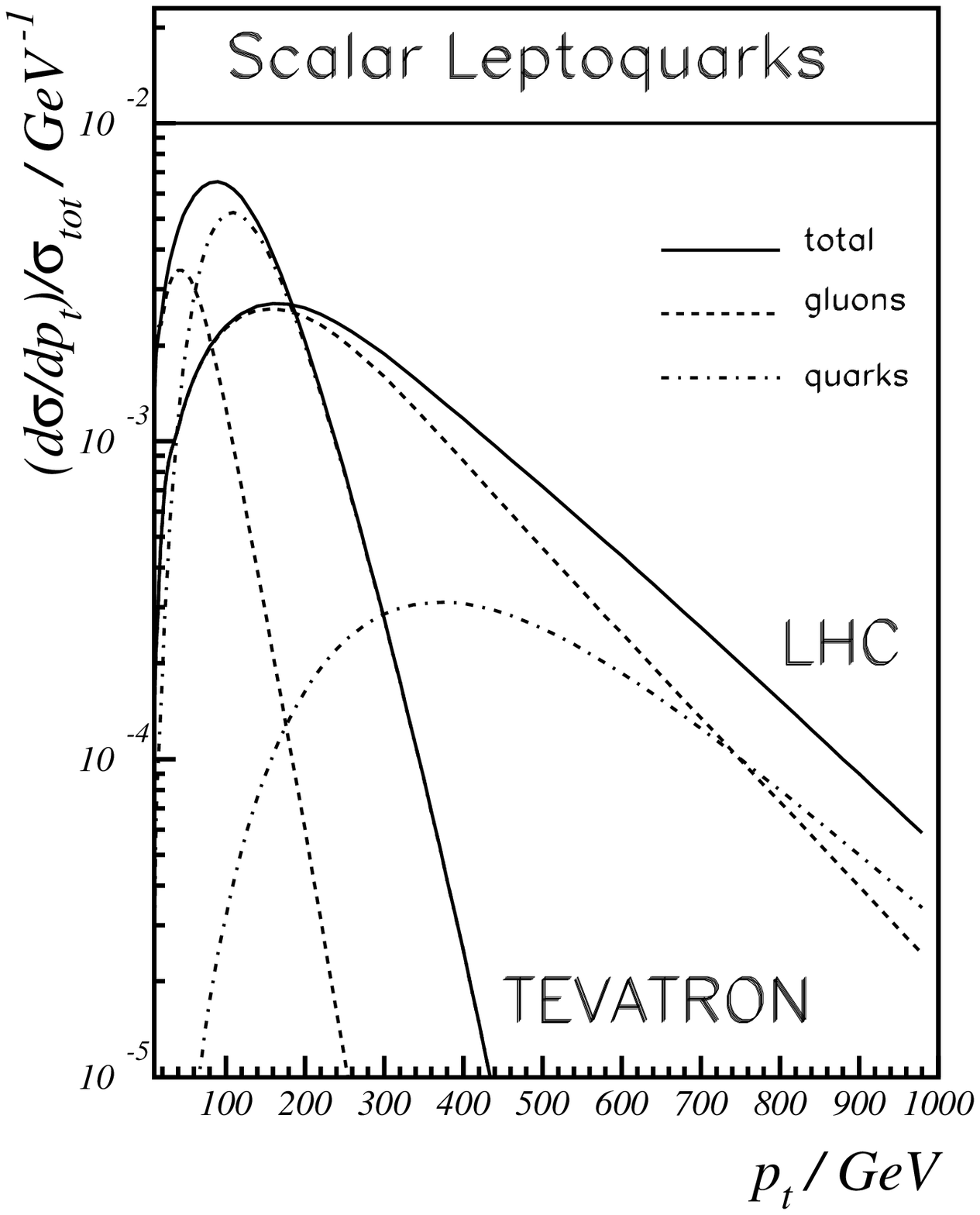,height=18cm,width=16cm}}

\vspace{2mm}
\noindent
\small
\end{center}
{\sf
Figure~8a:~$p_{\perp}$ distributions for scalar leptoquark pair production
at the
TEVATRON and LHC.
The leptoquark masses are the same as in
figures~6a,b.}
\normalsize
\newpage
\begin{center}

\mbox{\epsfig{file=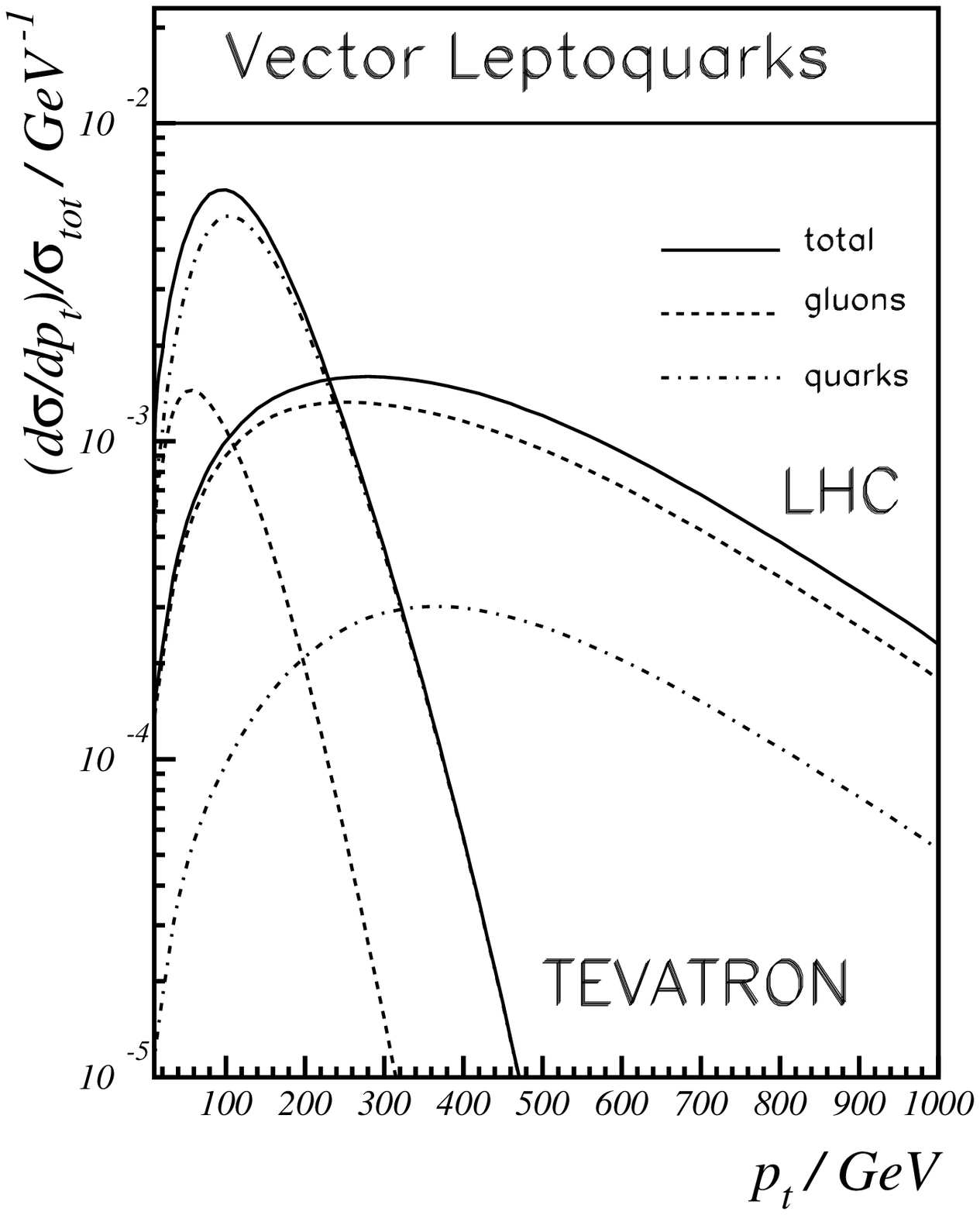,height=18cm,width=16cm}}

\vspace{2mm}
\noindent
\small
\end{center}
{\sf
Figure~8b:~$p_{\perp}$ distributions for vector leptoquark pair production
at the
TEVATRON and LHC assuming minimal vector couplings.
The notation is the same as in figure~8a.
}
\normalsize
\newpage
\begin{center}

\mbox{\epsfig{file=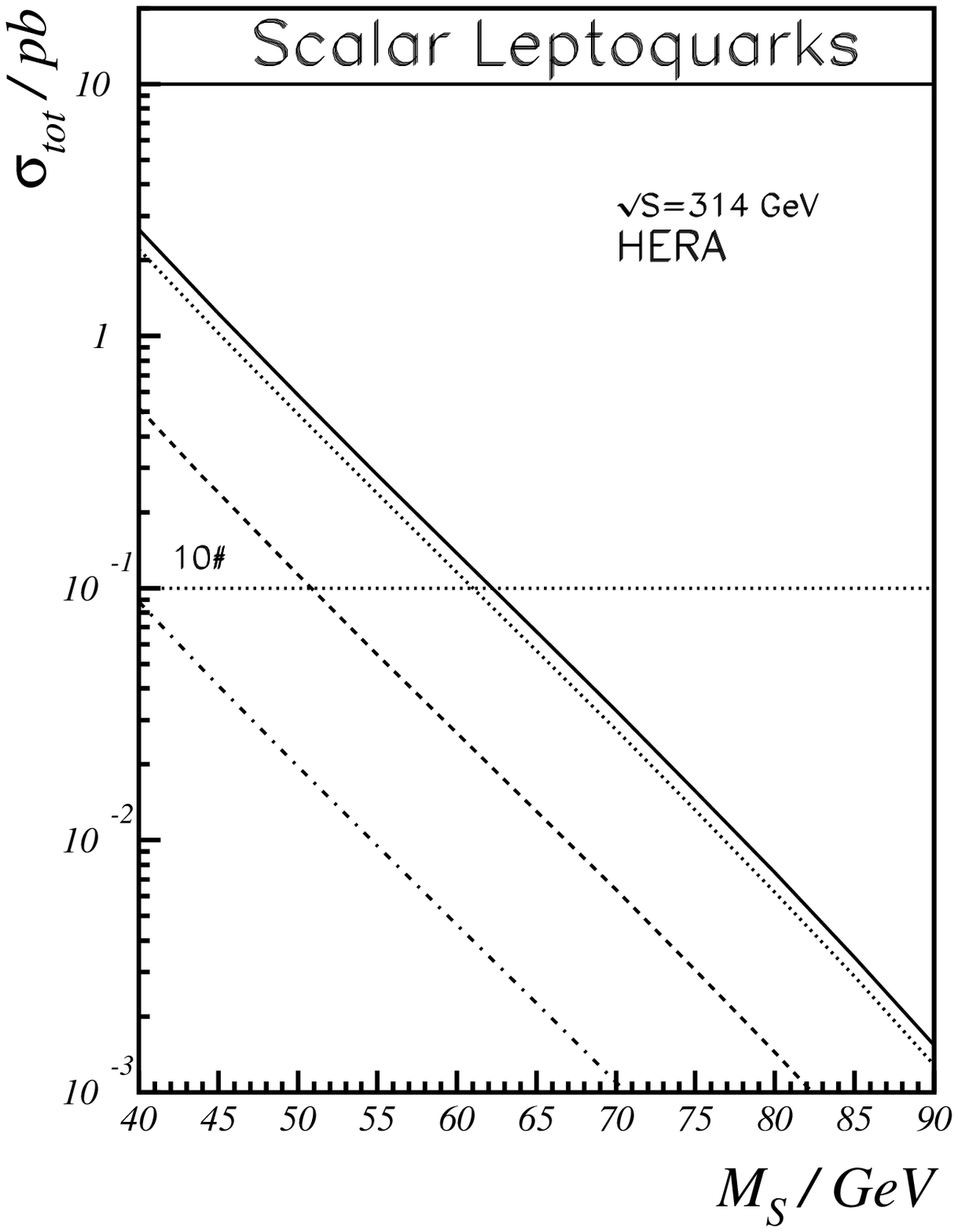,height=18cm,width=16cm}}

\vspace{2mm}
\noindent
\small
\end{center}
{\sf
Figure~9a:~Integrated cross
sections for scalar leptoquark pair production at HERA,
$\sqrt{S}~=~314~\GeV$.
Full line:~$\sigma_{tot}$ for $|Q_{\Phi}| = 5/3$;
dotted line:~$\sigma_{dir}$ for $|Q_{\Phi}| = 5/3$;
dashed line:~$\sigma_{tot}$ for $|Q_{\Phi}| = 1/3$;
dash--dotted line:~$\sigma_{dir}$ for $|Q_{\Phi}| = 1/3$.
}
\normalsize
\newpage
\begin{center}

\mbox{\epsfig{file=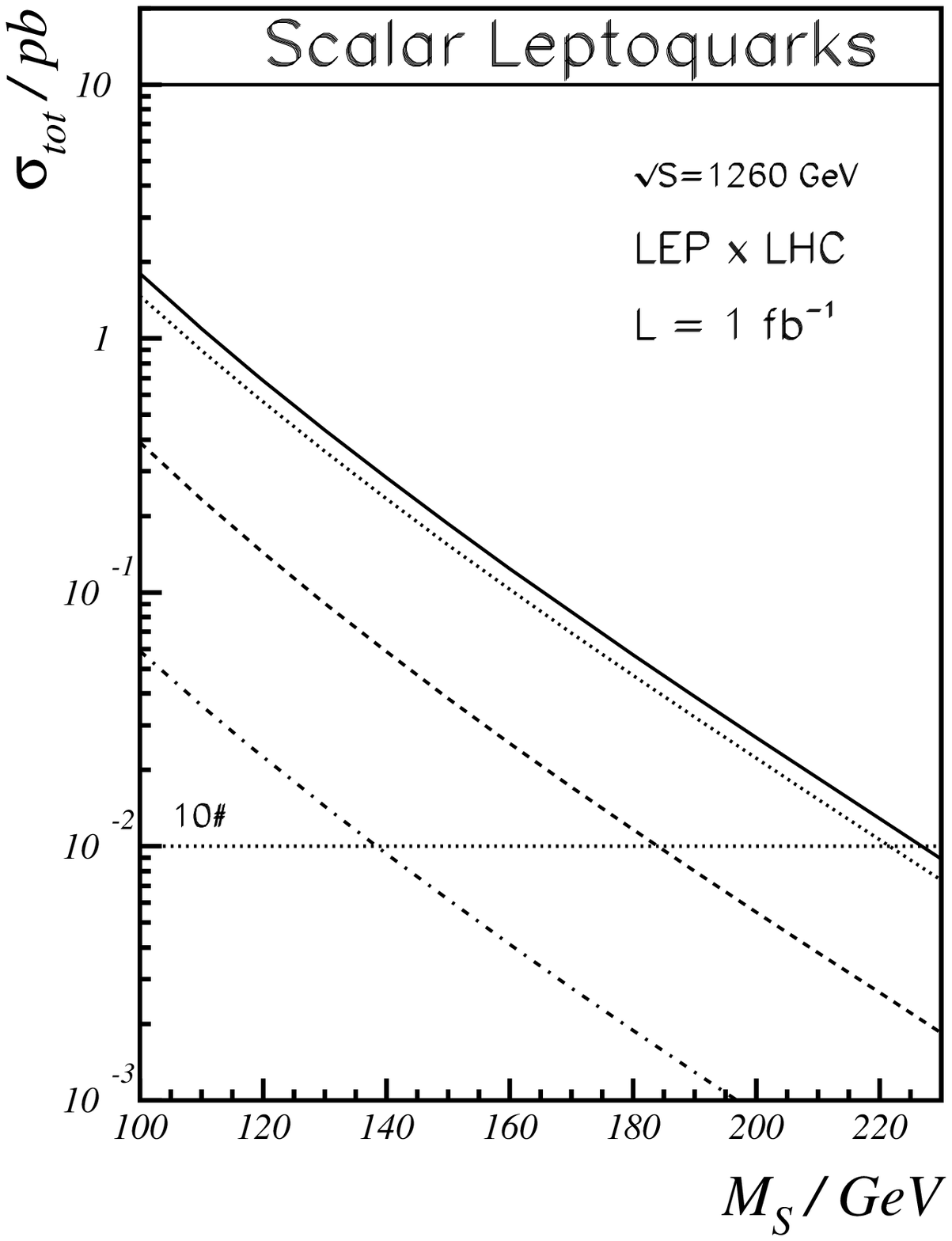,height=18cm,width=16cm}}

\vspace{2mm}
\noindent
\small
\end{center}
{\sf
Figure~9b:~Integrated cross
sections for scalar leptoquark pair production at LEP~$\otimes$~LHC,
$\sqrt{S}~=~1260~\GeV$. The notations are
the same as in figure~8a.
}
\normalsize
\newpage
\begin{center}

\mbox{\epsfig{file=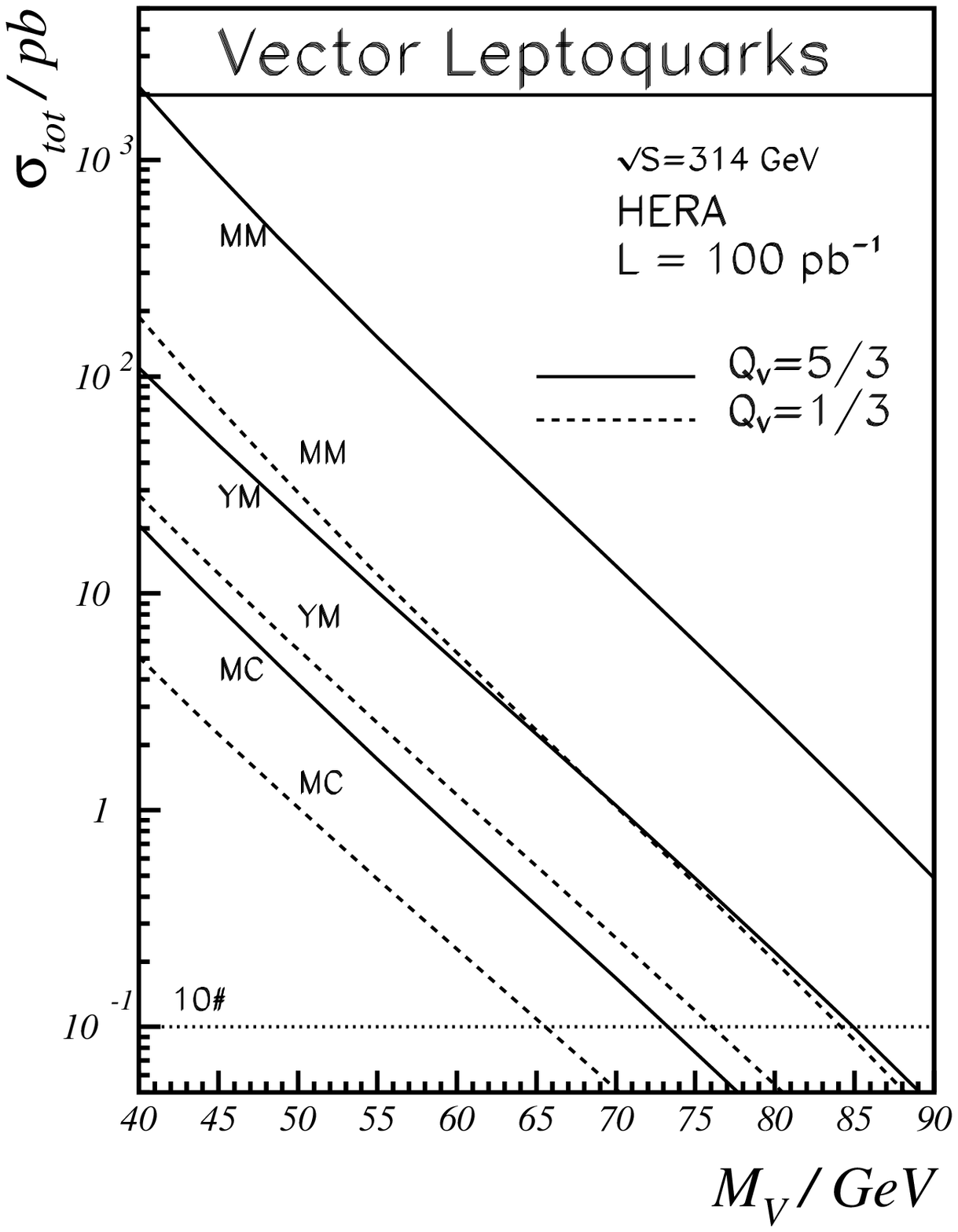,height=18cm,width=16cm}}

\vspace{2mm}
\noindent
\small
\end{center}
{\sf
Figure~10a:~Integrated cross
sections $\sigma_{tot} = \sigma_{dir} + \sigma_{res}$
for vector leptoquark pair production at HERA,
$\sqrt{S}~=~314~\GeV$.
The different choices for the anomalous couplings are:
$\kappa_{G,\gamma}=\lambda_{G,\gamma} = -1$ (MM);
$\kappa_{G,\gamma}=\lambda_{G,\gamma} =  0$ (YM);
$\kappa_{G,\gamma}=1, \lambda_{G,\gamma} =  0$ (MC).
}
\normalsize
\newpage
\begin{center}

\mbox{\epsfig{file=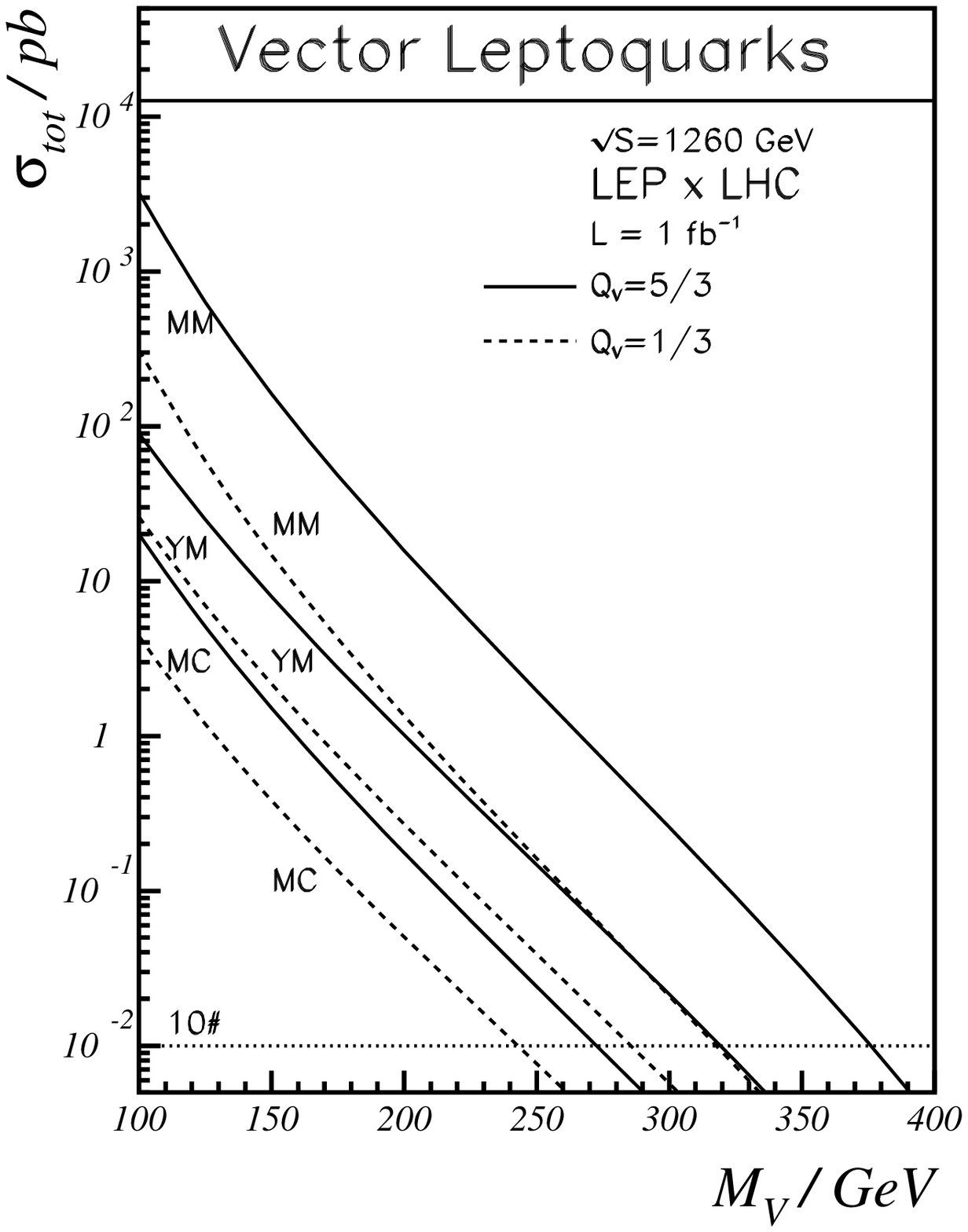,height=18cm,width=16cm}}

\vspace{2mm}
\noindent
\small
\end{center}
{\sf
Figure~10b:~Integrated cross
sections for vector leptoquark pair production at LEP~$\otimes$~LHC,
$\sqrt{S}~=~1260~\GeV$. The notations are
the same as in figure~9a.
}
\normalsize
\newpage
\begin{center}

\mbox{\epsfig{file=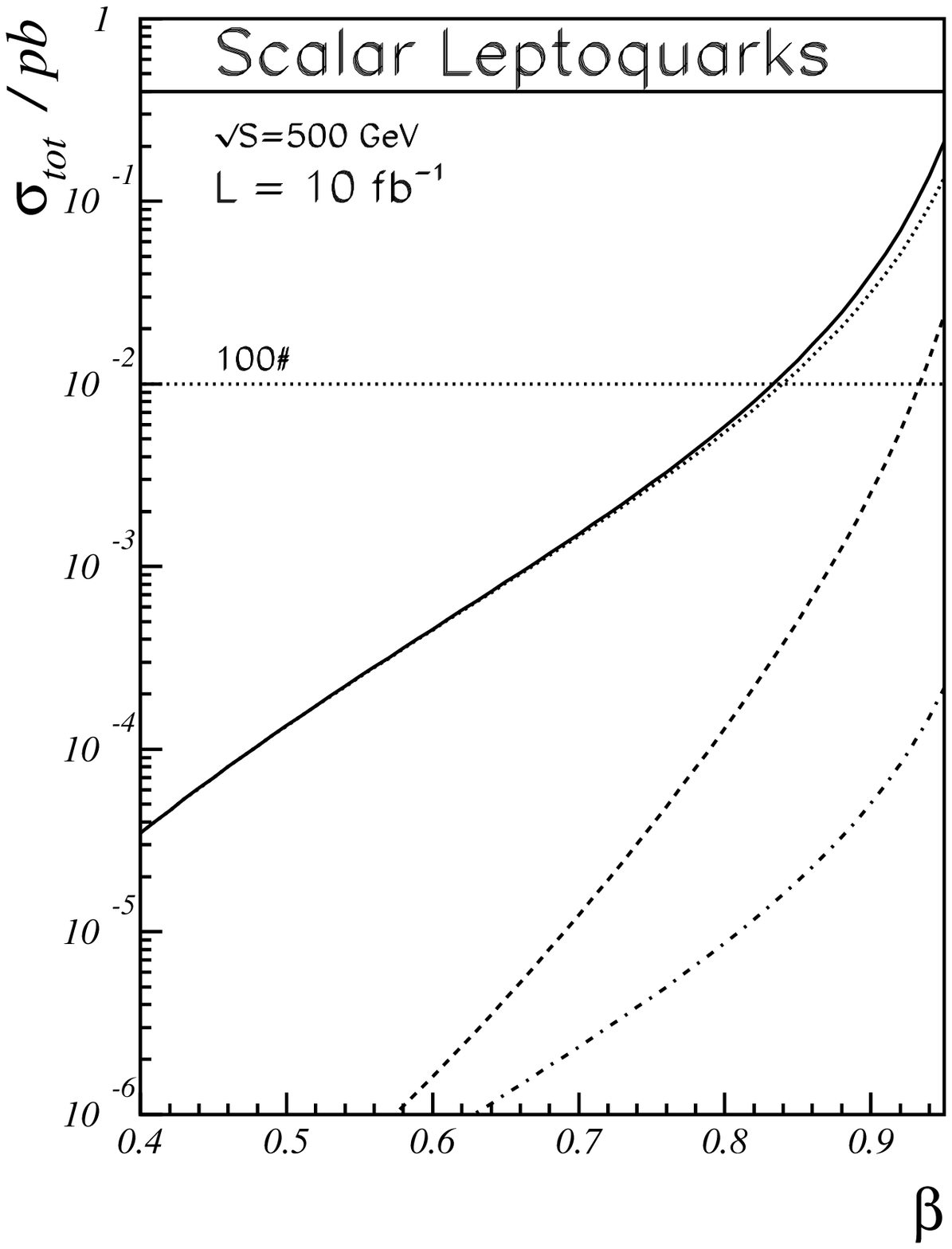,height=18cm,width=16cm}}

\vspace{2mm}
\noindent
\small
\end{center}
{\sf
Figure~11a:~Integrated cross
sections for scalar leptoquark pair production through
$\gamma^* \gamma^* \rightarrow S\overline{S}$ (WWA spectrum)
at future $e^+e^-$
colliders for
$\sqrt{S}~=~500~\GeV$ as a function of $\beta$.
Full line:~$\sigma_{tot}$ for $|Q_{\Phi}| = 5/3$;
dotted line:~$\sigma_{dir}$ for $|Q_{\Phi}| = 5/3$;
dashed line:~$\sigma_{tot}$ for $|Q_{\Phi}| = 1/3$;
dash--dotted line:~$\sigma_{dir}$ for $|Q_{\Phi}| = 1/3$.
}
\normalsize
\newpage
\begin{center}

\mbox{\epsfig{file=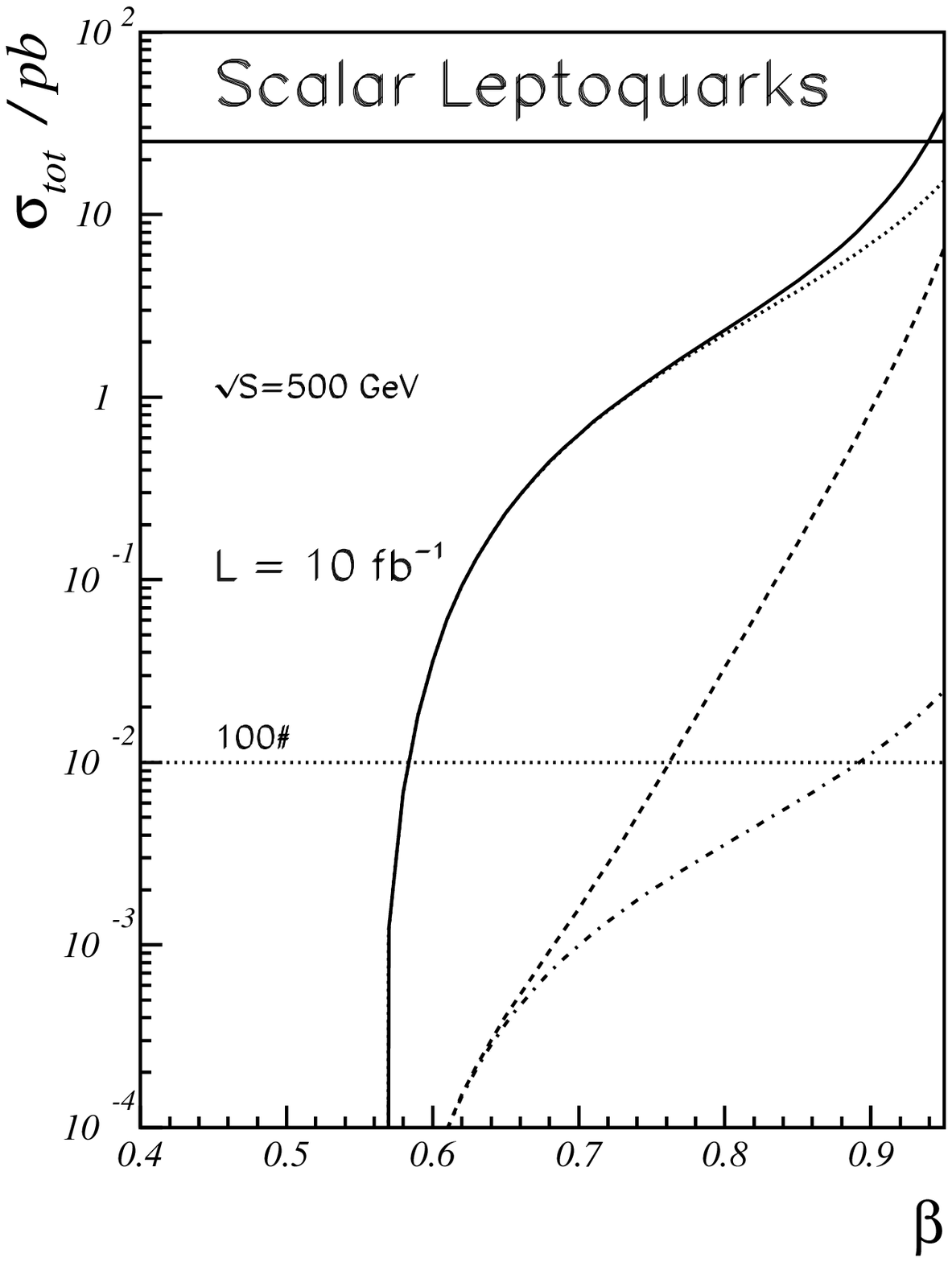,height=18cm,width=16cm}}

\vspace{2mm}
\noindent
\small
\end{center}
{\sf
Figure~11b:~Integrated cross
sections for scalar leptoquark pair production at future
$\gamma \gamma$ colliders using laser back scattering for electron
beam conversion. The parameters are the same as in figure~11a.
}
\normalsize
\newpage
\begin{center}

\mbox{\epsfig{file=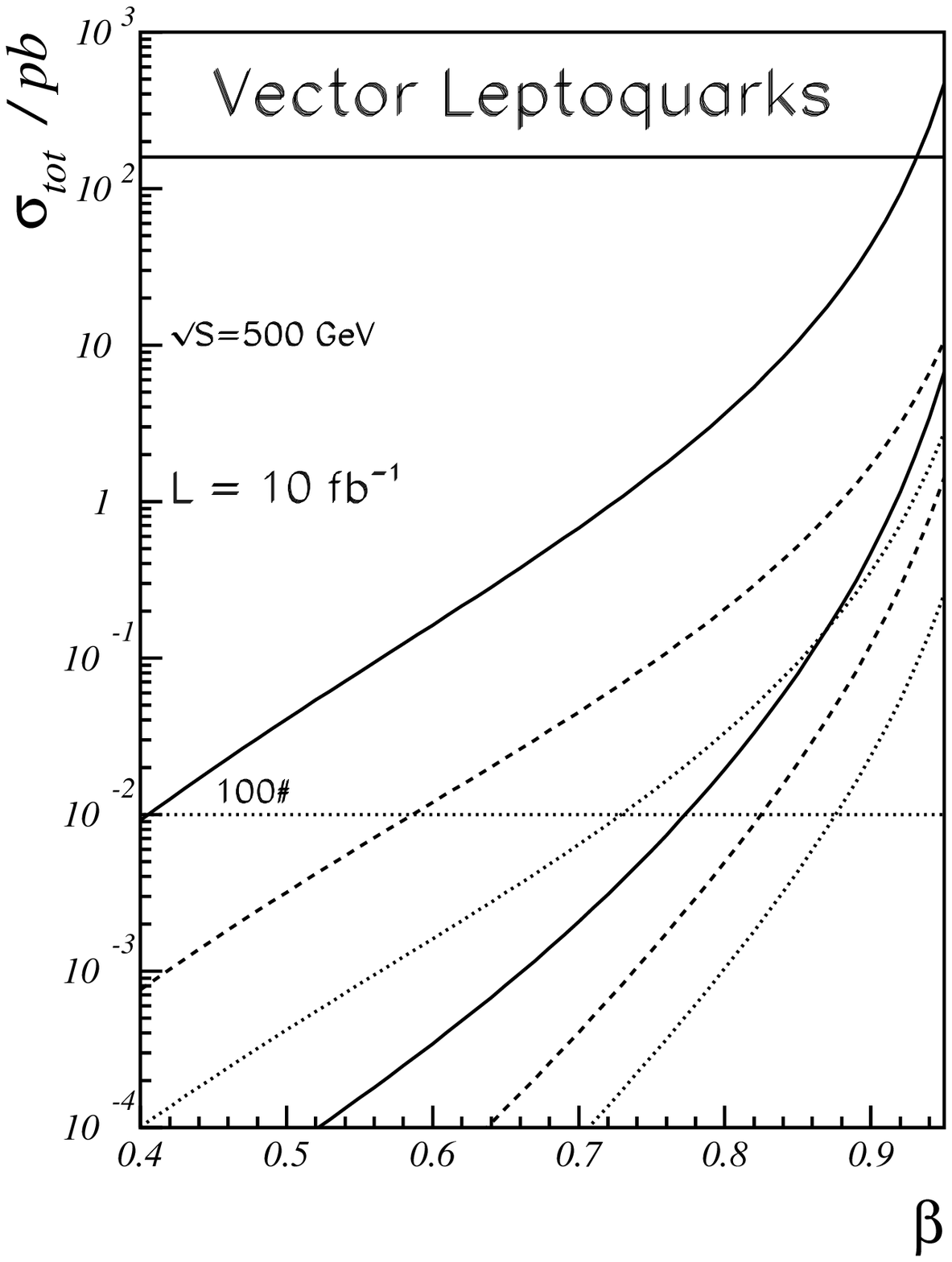,height=18cm,width=16cm}}

\vspace{2mm}
\noindent
\small
\end{center}
{\sf
Figure~12a:~
Integrated cross
sections for vector leptoquark pair production through
$\gamma^* \gamma^* \rightarrow V\overline{V}$ (WWA spectrum)
at future $e^+e^-$
colliders for
$\sqrt{S}~=~500~\GeV$ as a function of $\beta$.
Upper full line:~$|Q_{\Phi}| = 5/3, \kappa_{\gamma,G}
= \lambda_{\gamma,G} = -1$;
Upper dashed line:~$|Q_{\Phi}| = 5/3, \kappa_{\gamma,G}
= \lambda_{\gamma,G} = 0$;
Upper dotted line:~$|Q_{\Phi}| = 5/3, \kappa_{A,G} = 1,
\lambda_{\gamma,G} = 0$; The corresponding lower lines represent
for $|Q_{\Phi}| = 1/3$.
}
\normalsize
\newpage
\begin{center}

\mbox{\epsfig{file=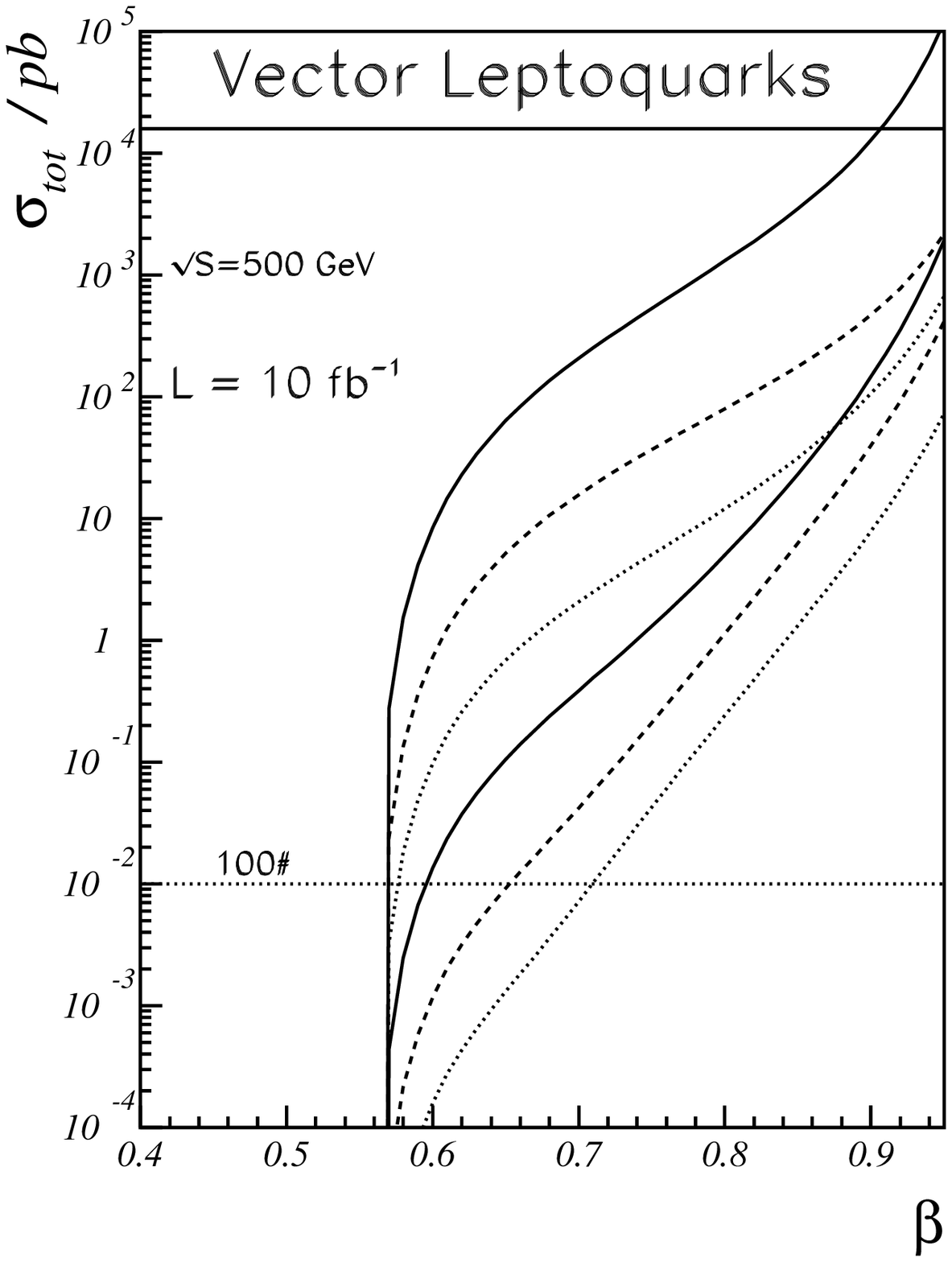,height=18cm,width=16cm}}

\vspace{2mm}
\noindent
\small
\end{center}
{\sf
Figure~12b:~Integrated cross
sections for vector leptoquark pair production at future
$\gamma \gamma$ colliders using laser back scattering for electron
beam conversion. The parameters are the same as in figure~12a.
}
\normalsize
\end{document}